\documentclass[acmsmall]{acmart}
\usepackage[utf8]{inputenc}
\usepackage{graphicx}
\usepackage{todonotes}
\usepackage[table]{xcolor}
\usepackage{tcolorbox}
\usepackage{subcaption}

\usepackage{wrapfig}
\usepackage{footmisc}
\usepackage{tabu}
\usepackage{algorithm}
\usepackage{algpseudocode}
\usepackage{multirow}
\usepackage{adjustbox}
\usepackage{amsthm}
\usepackage{amsmath,amsfonts}
\usepackage{hhline}
\usepackage{url}
\PassOptionsToPackage{normalem}{ulem}
\usepackage{ulem}
\usepackage{array}
\usepackage{type1cm}
\usepackage[T1]{fontenc}
\usepackage{microtype}
\usepackage{fancyhdr}
\algnewcommand\algorithmicforeach{\textbf{for each}}
\algnewcommand{\LineComment}[1]{\(\triangleright\) #1}
\algdef{S}[FOR]{ForEach}[1]{\algorithmicforeach\ #1\ \algorithmicdo}

\definecolor{darkgreen}{RGB}{21,176,26}
\definecolor{orange}{RGB}{242,148,7}

\newcommand{\hard}{\mathcal{H}\mathrm{ard}}
\newcommand{\chard}{C^{(\hard)}}
\newcommand{\opp}{\mathcal{O}\mathrm{pp}}
\newcommand{\avgopp}{\overline{\mathcal{O}\mathrm{pp}}}
\newcommand{\copp}{C^{(\opp)}}

\newcommand{\iewithinp}{IEWithin\_p}
\newcommand{\ieacrossp}{IEAcross\_p}
\newcommand{\note}[1]{\textbf{Note.}\ #1}

\newcommand{\acc}{\text{Acc}}

\graphicspath{{figures/}}

\setcopyright{none}

\begin{document}

\title{Hardness, Structural Knowledge, and Opportunity: An Analytical Framework for Modular Performance Modeling}

\pagestyle{fancy}

\fancyhead{}  

\fancyhead[C]{\scriptsize Hardness, Structural Knowledge, and Opportunity: An Analytical Framework \hfill \thepage}  


\author{Omid Gheibi}
\affiliation{
	\institution{Carnegie Mellon University}
	\city{Pittsburgh, Pennsylvania}
	\country{USA}
}
\email{omid.gheibi@gmail.com}
 
\author{Christian Kästner}
\affiliation{
	\institution{Carnegie Mellon University}
	\city{Pittsburgh, Pennsylvania}
	\country{USA}
}
\email{kaestner@cs.cmu.edu}

\author{Pooyan Jamshidi}
\affiliation{
	\institution{University of South Carolina}
	\city{Columbia, SC}
	\country{USA}
}
\email{pjamshid@cse.sc.edu}

\begin{abstract}
Performance-influence models are beneficial for understanding how configurations affect system performance, but their creation is challenging due to the exponential growth of configuration spaces.
While gray-box approaches leverage selective \emph{structural knowledge} (like the module execution graph of the system) to improve modeling, the relationship between this knowledge, a system's characteristics (we call them \emph{structural aspects}), and potential model improvements is not well understood.
This paper addresses this gap by formally investigating how variations in structural aspects (e.g., the number of modules and options per module) and the level of structural knowledge impact the creation of \emph{opportunities} for improved \emph{modular performance modeling}.
We introduce and quantify the concept of modeling \emph{hardness}, defined as the inherent difficulty of performance modeling.
Through controlled experiments with synthetic system models, we establish an \emph{analytical matrix} to measure these concepts. Our findings show that modeling hardness is primarily driven by the number of modules and configuration options per module.
More importantly, we demonstrate that both higher levels of structural knowledge and increased modeling hardness significantly enhance the opportunity for improvement.
The impact of these factors varies by performance metric; for ranking accuracy (e.g., in debugging task), structural knowledge is more dominant, while for prediction accuracy (e.g., in resource management task), hardness plays a stronger role.
These results provide actionable insights for system designers, guiding them to strategically allocate time and select appropriate modeling approaches based on a system's characteristics and a given task's objectives.
\end{abstract}

\ccsdesc[500]{Software and its engineering~Software performance}
\ccsdesc[400]{Computing methodologies~Machine learning}
\ccsdesc[400]{Computing methodologies~Causal reasoning and inference}

\keywords{Performance-influence modeling, Highly configurable systems, Modular systems, Machine learning, Causal inference, Structural knowledge}

\maketitle

\section{Introduction}
\label{sec: intro}

Performance-influence models help users understand how various configurations of option variables and their interactions affect system performance~\cite{kolesnikov2019tradeoffs, siegmund2015performance}. 
From a practical point of view, it can be useful for the debugging and resource management of the system by predicting behavior under diverse configurations, balancing accuracy, interpretability, and development cost~\cite{grebhahn2019predicting, guo2013variability, kaltenecker2020interplay, siegmund2015performance, jamshidi2017transfer,chen2020understanding, vitui2021mlasp, xu2016early}. 
For instance, a Memcached configuration option called \texttt{maxconns\_fast} is designed to drop new connections once the maximum limit is exceeded, returning an error.
A developer might set this option to \texttt{false} to reduce the rate of errors and improve user satisfaction.
However, this change could unexpectedly increase system costs by allowing too many connections to open, consuming a significant amount of memory.
In this scenario, a performance model (such as a regression formula) can help the developer debug the issue by identifying the precise impact of this configuration change on performance measures like memory usage and cost.
This allows them to pinpoint which configuration setting (in this case, setting \texttt{maxconns\_fast} to \texttt{false}) is causing the increase in system cost.

A central challenge in this domain lies in the sheer scale of potential configurations.
For example, a system with 90 binary options would have $2^{90}$ possible configurations.
To put this number into perspective, it's an unimaginably large quantity.
It is s an order of magnitude more than the number of grains of sand on all the beaches of Earth, or billions of times longer than the age of the universe in seconds.
This immense size makes it practically impossible to test every single configuration.
Thus, instead of brute-force testing, we need smarter, more efficient ways to understand how these configurations influence performance.

To that end, three primary approaches are commonly used for system analysis and performance modeling: black-box, white-box, and gray-box. The black-box approach treats the system as an opaque entity, relying on a set of training data (i.e., sampled configurations and their corresponding performance measurements) and machine learning methods (e.g., regression methods) to model their relationship~\cite{grebhahn2019predicting, guo2013variability, kaltenecker2019distance, kaltenecker2020interplay, sarkar2015cost, siegmund2012predicting, siegmund2012spl, siegmund2015performance}. 
The white-box approach uses detailed internal knowledge, such as source code analysis, to understand configuration impacts, offering greater interpretability but facing challenges with complex systems, e.g., scalability and limited code access~\cite{velez2020configcrusher, velez2021white}. 
The gray-box approach combines external observations with selective \emph{structural knowledge}, such as key metrics and system structures like the logical boundary of modules, making it ideal for restricted-access scenarios like distributed systems~\cite{chao2018gray, arpaci2001information, liang2023modular}.

Yet, in the gray-box approach to performance modeling of distributed systems, such as those designed with microservices, the extent to which \emph{structural knowledge} contributes to an effective performance model remains poorly understood. 
Insufficient knowledge may lead to missed opportunities for improving performance-influence models. 
Conversely, acquiring additional knowledge can be costly and may degrade the performance of existing methods by necessitating extra steps to identify the most relevant information or requiring adjustments to the modeling approach that demand more configuration measurements. 
Furthermore, the impact of structural knowledge on developing an effective performance model may depend on architectural characteristics of the distributed system, such as the number of modules and their respective configuration options.
In this paper, we call these architectural characteristics \emph{structural aspects}.

\note{We distinguishes between the two concepts: \emph{structural knowledge} is the information used to improve modeling (like a module execution graph), while \emph{structural aspects} are the system's own characteristics (like the number of modules) that affect modeling.
	We investigates the impact of these two concepts on performance modeling.}

This article aims to address this gap in understanding.
Therefore, a more formal goal definition of this research is:

\begin{quote}
	\emph{
		\textbf{Goal}.
		Exploring the impact of variations in structural aspects and levels of structural knowledge on how well modular performance modeling works in distributed systems.
	}
\end{quote}

The potential for improving the efficacy of performance modeling, which we refer to as \emph{opportunity}, is inherently linked to the \emph{hardness} of the modeling problem, i.e., the intrinsic difficulty of the problem in the first place, and is mediated by structural knowledge.
Hardness itself stems from the system's structural aspects (e.g., the number of modules and interactions between configuration options), but it is the availability of structural knowledge that can make the modeling task easier, and creating an opportunity for improvement.
For instance, consider two modular software systems, each with a configuration space size of $2^{90}$.
A system with many modules and complex interactions between configuration options seems to be harder to model than one with fewer modules and interactions, as the former might require significantly more measured configurations (i.e., training data for machine learning methods) to achieve comparable efficacy.\footnote{
	Systems with higher complexity generally require models with more parameters to achieve comparable efficacy, thereby increasing the need for higher number of measurements in the configuration space (e.g.,  \cite{vapnik2013nature, gheibi21impact} studied this impact, particularly in terms of accuracy measures).
	This relationship can be understood through the concept of the VC-dimension (Vapnik–Chervonenkis dimension), which quantifies the capacity of a learning model. A higher number of parameters increases the VC-dimension, and the VC-dimension is probabilistically related to the accuracy of the model. 
	Consequently, a system with greater structural complexity requires more measured configurations to attain a specific accuracy level.
}

Accordingly, the motivation for this study is to address the gap in understanding how structural knowledge of a modular system can improve performance modeling. While some research suggests that structural knowledge can be leveraged for better performance modeling, a systematic investigation into how varying levels of this knowledge affect the performance modeling efficacy is lacking. 
As an initial phase of investigation, we used a simulated system called the ``Self-care Mobile App''. 
Initial experiments with this app showed that incorporating structural knowledge did not improve performance modeling efficacy compared to a black-box approach. However, when the same experiment was conducted on a real-world application, the "Hotel Reservation" app, leveraging structural knowledge did significantly improve the model's efficacy. 
This discrepancy, which we explained in detail in Section~\ref{sec: problem of hardness and opportunity}, led to the central research questions:

\begin{tcolorbox}[colback=gray!20, 
	colframe=gray!50, 
	boxrule=0.5mm, 
	arc=4mm, 
	width=\linewidth, 
	title={RQ1: How do changes in structural aspects of modular systems (Section~\ref{sec: structural aspects}) affect the hardness of performance modeling (Section~\ref{sec: hardness})?},
	fonttitle=\bfseries, 
	sharp corners=northwest] 
	
	\textbf{Short Answer}.
	The primary structural factors influencing modeling hardness are the number of modules and the number of options per module.
\end{tcolorbox}

\begin{tcolorbox}[colback=gray!20, 
	colframe=gray!50, 
	boxrule=0.5mm, 
	arc=4mm, 
	width=\linewidth, 
	title={RQ2: How do the levels of structural knowledge (Section~\ref{sec: structural knowledge}) and the modeling hardness impact the creation of opportunity (Section~\ref{sec: opportunity}) for improving modular performance modeling?},
	fonttitle=\bfseries, 
	sharp corners=northwest] 
	
	\textbf{Short Answer}. Higher structural knowledge is more effective for improving ranking accuracy (used for tasks like debugging), while increased modeling hardness has a stronger impact on creating opportunity for prediction accuracy (used for tasks like resource management).
\end{tcolorbox}

\paragraph{Method Overview} 
To address these research questions, we employed analytical statistical methods to formulate and evaluate corresponding hypotheses. 
We conducted controlled experiments using simulated system models, which allowed us to manipulate independent variables. 
This approach was necessary because a review of public repositories revealed a lack of real-world applications with the required structural diversity~\cite{meiklejohn2021service}.
In these controlled experiments, the independent variables are the structural aspects of the system and the level of structural knowledge. 
The dependent variables, which are influenced by the independent variables, are the hardness of performance-influence modeling and the opportunity for its improvement.

\paragraph{Practical Implications} 
This research provides a framework for system designers to make informed decisions. To achieve cost-effective performance modeling, a designer can use the following insights:
\begin{itemize}
	\item For high-hardness systems (e.g., those with a large number of modules and options per module), the priority should be to acquire more detailed structural knowledge.
	This is particularly beneficial for high-stakes tasks like debugging, where high-accuracy ranking is beneficial.
	
	\item For low-hardness systems, a less detailed level of structural knowledge may be sufficient to achieve a desired performance model.
	This applies to tasks where a general prediction is needed, such as planning for resource management.
\end{itemize}

\noindent
By understanding the distinct impacts of structural knowledge on ranking versus prediction accuracy, designers can select the most appropriate modeling approach to match their specific objectives, thereby allocating time and effort more efficiently.

\paragraph{Limitations} 
This study's findings are based on controlled experiments with synthetic system models, which were necessary to isolate and quantify the concepts of hardness and opportunity. Due to the limited number of real-world distributed applications with the required structural diversity for obtaining strong statistical results, this approach was a necessary trade-off. As such, a primary limitation is the potential for these results to not generalize perfectly to all real-world software systems. While we designed the synthetic models to be representative of modular systems, real-world complexity, including hidden dependencies and unexpected emergent behaviors, may introduce factors not captured in our experiments.

\paragraph{Contributions.} 
Main contributions of this paper are as follows:
\begin{itemize}
	\item \emph{Hardness and Opportunity Analytical Matrix}: This paper introduces a new analytical matrix to understand how the structural aspects of a modular system, such as the number of modules and configuration options per module, influence the hardness of performance modeling. 
	This matrix also shows how different levels of structural knowledge can create how much opportunity to improve the performance models, which helps to guide the system designer in choosing an effective modeling approach.
	\item \emph{Factors Influencing Modeling Hardness}: This study finds that the number of modules and the number of configuration options per module are the primary drivers of modeling hardness. In contrast, the probability of influence relationships (i.e., causal edges) between and within modules have a negligible effect on hardness.
	\item \emph{Impact of Structural Knowledge on Opportunity}: (i) This paper demonstrates that higher levels of structural knowledge consistently lead to a greater opportunity for improving ranking accuracy (as measured by the Spearman Correlation Coefficient). This benefit is significant across all levels of system complexity. 
	(ii) The impact of structural knowledge on prediction accuracy (measured by the MAAPE metric) is more limited. Significant benefits are only observed in high-hardness scenarios (i.e., complex systems), suggesting that for simpler systems, less detailed knowledge is sufficient.
	\item \emph{Practical Guidance for System Designers}: The findings provide a clear guide for system designers on how to prioritize the acquisition of structural knowledge. For tasks like debugging, where ranking configurations is key, designers should invest in detailed knowledge (e.g., execution graphs). For performance tasks focused on prediction, detailed knowledge is most beneficial for highly complex systems.
\end{itemize}

\section{Related Work}
\label{sec: related work}
Configuration options in systems are deferred design decisions. 
Rather than having a one-size fits all system that more or less supports different workloads, configuration options allow a user to chose different decisions tailored to their workload or use case, making different tradeoff decisions. 
For example, rather than fixing a specific behavior, based on a configuration option the same video encode could produce less compact files very quickly or a smaller file more slowly.

Hence, configuration decisions can substantially influence performance and the tradeoffs with other qualities. In this context, developers may care about purely optimizing performance or about understanding the influence of options on performance (and other qualities). The latter is more important for debugging or to make informed tradeoff decisions than just pure optimization. Hence, different techniques are tailored for the different use cases of performance optimization and performance modeling.

This section organizes related research into two primary categories: performance optimization and performance modeling. 
It further divides performance modeling research into three subcategories: sampling, learning, and leveraging structural knowledge in performance modeling.

\paragraph{Performance Optimization} 
The first category of related work focuses on optimizing software system performance. 
Some studies in this area utilize performance models, employing search methods such as Bayesian optimization~\cite{wu2022autotuning, liu2021gptune, chen2021efficient, shao2024efficient, wu2025ytopt} or supervised machine learning methods~\cite{fursin2011milepost, zhu2025pdcat, leather2014automatic, sajjadinasab2024graph} to explore the parameter space. 
Other studies do not rely on a performance model, instead using local search techniques like hill-climbing based on apriori or kernel distributions~\cite{ansel2014opentuner, kulkarni2003finding, kisuki2000combined, pan2006fast}, or evolutionary algorithms~\cite{petke2017genetic,gao2025grouptuner,aragon2023automatic, ansel2012siblingrivalry}. However, this current study's primary focus is on performance modeling rather than direct performance optimization.

\paragraph{Performance Modeling (Sampling)}
In many practical scenarios, manually collecting data to build a performance-influence model is infeasible due to the exponential growth of the configuration space.
To address this challenge, researchers have proposed various sampling techniques that select a representative subset of configurations for measurement.
These techniques aim to reduce the number of measurements required while still achieving effective performance modeling, such as accurate prediction.
Research indicates that no single sampling scheme is universally optimal; its effectiveness depends on factors like configuration space structure and modeling objectives~\cite{grebhahn2019predicting, bessal2025unveiling, kaltenecker2020interplay, alves2020sampling}. 
Some studies address coverage and high-dimensionality challenges~\cite{bao2018autoconfig, kaltenecker2019distance, grebhahn2019predicting}, while others propose adaptive and progressive sampling methods~\cite{nair2017using, ha2019deepperf}. 
Further research has explored white-box analysis and feature selection~\cite{schmid2022performance, han2021confprof, schmid2024cost, velez2020configcrusher}, as well as workload sensitivity and cross-environment learning (e.g., using transfer or causal learning) to reduce sample sizes~\cite{martin2021transfer, muhlbauer2023analysing, jamshidi2017transfer, iqbal2022unicorn}. 
This study differentiates itself by focusing on understanding the impact of structural knowledge level on the efficacy of the performance model (i.e., machine learning models), unlike the aforementioned studies.
It also evaluates this impact by analyzing the sample size needed to achieve a target level of efficacy in performance modeling (similar to these studies), while keeping the sampling method constant. 

\paragraph{Performance Modeling (Learning)} 
While there is no single best universal learning method (same as the sampling method) for performance modeling~\cite{grebhahn2019predicting}, various machine learning approaches have been applied. 
These include classical and tree-based methods such as linear models, LASSO, random forests, and gradient boosting, which are commonly used~\cite{grebhahn2019predicting, valov2015empirical, ha2019performance}. 
A second set of studies has employed neural network methods, including deep learning~\cite{ha2019deepperf, ritter2021noise, shu2020perf, gong2024deep}. 
Thirdly, some research has utilized causal machine learning for performance modeling~\cite{iqbal2022unicorn, liang2023modular,iqbal2023cameo, chen2025promisetune}. 
Although this paper uses established learning approaches for performance modeling (i.e., random forest method from the tree-based approach, and the structural causal model from the causal learning approach), its main objective is not to propose a new learning method. 
Instead, it aims to understand how varying levels of structural knowledge impact the efficacy of modular performance modeling.

\paragraph{Leveraging Structural Knowledge In Performance Modeling} 
Most performance modeling approaches consider a system as a single unit, often relying on end-to-end measurements. However, some studies have explored leveraging the \emph{compositionality} of their methodology~\cite{freiling2006composition}. These studies, which often take a \emph{white-box approach}, utilize fine-grained structural knowledge such as code regions, call graph nodes, and traced functions to enhance modeling effectiveness~\cite{velez2021white, velez2020configcrusher, weber2021white, han2021confprof, chen2023diagconfig, meinicke2016essential, nguyen2016igen, reisner2010using}. 
However, many of these methods are very low-level and might not be scalable.

An alternative gray-box approach has also been proposed, which is particularly interesting because it offers a good balance of efficacy and scalability, and is the approach we build upon.
This approach does not deeply analyze code and its behavior. 
Instead, it involves inserting probes into the system to identify underlying structural knowledge, going beyond simple end-to-end measurements. 
These studies leverage intermediate variables and extract causal influences between these variables and configuration options to improve performance modeling efficacy~\cite{iqbal2023cameo, iqbal2022unicorn}. 
More recently, research in modular performance modeling has leveraged the execution graph of distributed systems and intermediate variables within each module, based on knowledge of module logical boundaries, to create scalable and effective performance models~\cite{liang2023modular}.

While these prior studies have successfully used structural knowledge, they have not investigated a critical gap in the field: the impact of varying levels of structural knowledge on the creation of opportunities to improve performance models. 
Furthermore, they have not explored how a system's structural aspects influences modeling hardness and the subsequent impact on opportunity creation. 
Addressing this gap is crucial for developing more scalable and effective performance models by providing a deeper understanding of the trade-offs between the level of structural knowledge, modeling hardness, and opportunity for improvement.

\paragraph{Sample Complexity.}
Another group of related studies falls within the field of theoretical machine learning.
These works primarily investigate the relationship between the number of training samples, the training error, and the generalization (or test) error of specific learning models, such as random forests and neural networks.
Some of these studies, like those by \cite{grunwald2019tight}, \cite{gonen2018average}, and \cite{puchkin2023exploring}, rely on specific data distribution assumptions, while others, such as \cite{shamir2015sample}, \cite{koltchinskii2000rademacher}, and \cite{tsybakov2003optimal}, are distribution-agnostic. While these studies offer a theoretical foundation for the practical concept of hardness we introduce, they do not consider the practical implications of their findings.
Specifically, they do not account for the impact of a system's structural aspects on this hardness, nor do they explore its interaction with the level of structural knowledge to enhance the learning model's efficacy, all of which are central to our work.

\section{Motivation}
\label{sec: problem of hardness and opportunity}

This section illustrates the motivation behind the research questions and the approach taken to address them in this paper. 
The motivation for this study stems from studies like~\cite{liang2023modular} that structural knowledge of modular software systems can be leveraged to improve the quality of performance modeling. 
However, there is a recognized scarcity of research in this area, particularly regarding the systematic use of different levels of structural knowledge to enhance performance modeling. 
This leads to the central question of how varying levels of structural knowledge impact the improvement of performance modeling in modular software systems.

To answer this question, a statistical analysis on a large number of modular software systems with a sufficient variety of structures was necessary, including variations in the number of configurations, modules, and interactions both within and across modules. 
A review of public repositories revealed a lack of real-world applications with the necessary structural diversity to conduct such a statistical analysis, a finding supported by the work of~\cite{meiklejohn2021service}.

\subsection{A First Failed Experiment: Where Structural Knowledge Did Not Enhance Modeling}

Excited by the proposal of recent structural and modular performance modeling approaches (see Section~\ref{sec: related work}), we tried to apply them ourselves. We conducted an initial experiment using a simulated, distributed software system we called the ``Self-care Mobile App''.\footnote{The Self-care Mobile App, which is inspired by the service-based model proposed in~\cite{quin2022reducing}, is an instance of the synthetic modular software model proposed in Section~\ref{subsec: Synthtetic Dataset Creation} for creating the synthetic dataset in this paper.}
This application consists of six services, each with six identical instances deployed on a multi-cloud system. The incoming traffic to each service is uniformly distributed among these servers if the corresponding option variables are configured as true for that service. 
Each service has six binary options to control the activation of these instances, resulting in a total configuration space of $2^{36}$. 
The performance of this application is measured by a utility function that combines the system's total cost and response time.

\begin{figure}[h]
	\centering
	\includegraphics[scale=0.5]{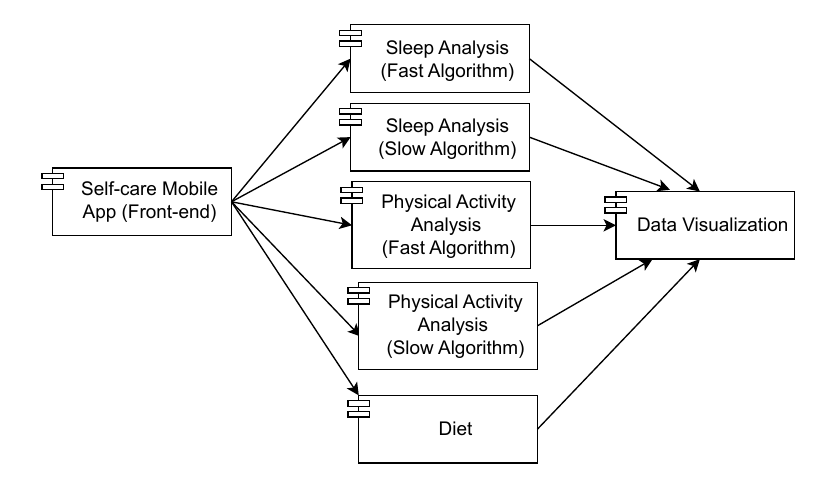}
	\caption{Execution graph of a synthetic modular and distributed software system, called Self-care Mobile App (inspired by the proposed service-based system  in~\cite{quin2022reducing}), developed based on the model explained in Section~\ref{subsec: Synthtetic Dataset Creation} for creating the synthetic dataset.}
	\label{fig: self-care execution graph}
\end{figure}

The initial task was to debug the application's performance by identifying which configuration options have the most significant impact on system utility under a fixed workload. 
A common challenge for developers is tuning complex applications with numerous configuration options, often without a complete understanding of how each option, or combination of options, affects performance. 
For this, we first employed a black-box approach, as it is a common starting point in performance analysis. 
This method treats the system as a monolithic entity and uses a regressor, such as the commonly used random forest, to model performance based on a sample of configurations.
Because evaluating all $2^{36}$ configurations is impractical, the model was trained on a subset of configurations to generate feature importance scores. 
These overall scores are particularly useful in a practical setting where developers need to efficiently prioritize their debugging and tuning efforts. 
However, the model can also be used to predict the precise influence of an option given a specific configuration, offering more granular insight into its impact. 
This is crucial because a change made to improve one aspect of performance, such as activating a server for a specific service to improve response time, can have a direct tradeoff, such as increasing the overall cost. Furthermore, while developers may have an intuition about some options, their expectations might be wrong, and they may be unaware of critical interactions. The random forest model is especially powerful here, as it can model and reveal these complex relationships between multiple parameters, which would be difficult to measure in isolation.

Additionally, we conducted these experiments using a range of training data sizes, from 20 to 1000. 
A higher number of training data sizes incurs a higher cost of measurements, and we aimed to monitor the trade-off between any potential improvement in model efficacy and its corresponding measurement cost.
This range is a common and practical range used in the literature, as the number of required training data is usually linearly proportional to the logarithmic factor of the size of the configuration space of the corresponding system.

However, the black-box method does not account for the modular structure of the system, such as logical module boundaries and their associated intermediate variables. To address this, multiple levels of structural knowledge were introduced to assess their impact on improving the efficacy of black-box performance modeling.

Two levels of structural knowledge were examined for the Self-care application (a more complete and formalized version of these levels will be introduced in Section~\ref{sec: structural knowledge}):

\begin{itemize}
	\item[-] \emph{Partial}: This level includes knowledge of module boundaries, configuration options, and intermediate variables (e.g., CPU and memory usage per service). This represents a minimal but realistic baseline of knowledge commonly available in modular system design.
	\item[-] \emph{Practical}: This level extends the Partial knowledge by incorporating the system's execution graph, which captures interactions between modules. 
	This provides a more detailed, yet practically attainable, understanding of the system's structure.
\end{itemize}

These two levels were used to model performance, alongside a black-box (\emph{Null}) approach and an impractical \emph{Ideal} scenario. 
The \emph{Ideal} scenario assumes perfect, comprehensive knowledge of all internal system workings, comprising all intermediate variables. 
While such complete information is unattainable in real-world systems, this approach serves as a benchmark to establish the theoretical maximum performance achievable through structural knowledge. 

For the \emph{Null} and \emph{Ideal} levels, a classical random forest regressor was used.
For the \emph{Partial} level, a hierarchical chain of random forest regressors was employed to leverage the logical boundaries of modules.
A structural causal model was used for the \emph{Practical} level, which utilized random forest regressors to model more detailed internal interactions between variables. 
Note that a random forest regressor was used as the primary method for all levels to ensure the results could be compared effectively. 

\begin{figure}[!htbp]
	\centering
	\begin{subfigure}{0.49\textwidth} 
		\centering
		\includegraphics[width=\linewidth]{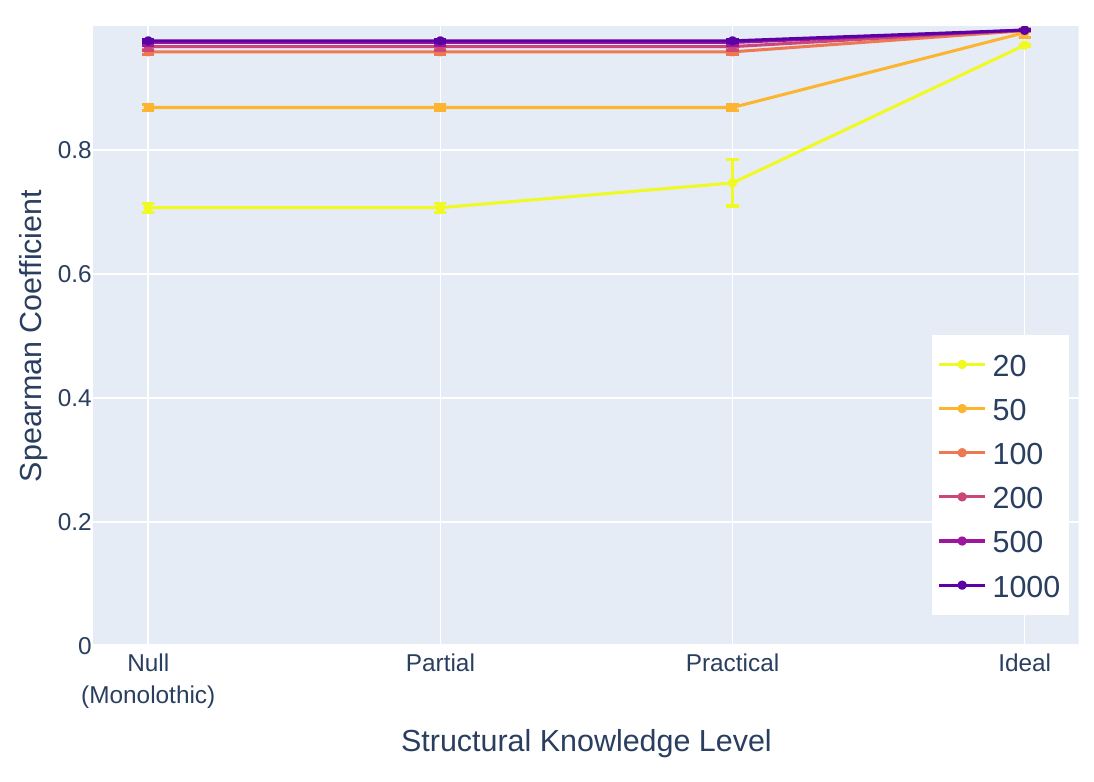}
		\caption{Self-care Mobile App}
		\label{fig: Spearman Coefficnet of All For Self-care}
	\end{subfigure}
	\hfill
	\begin{subfigure}{0.49\textwidth} 
		\centering
		\includegraphics[width=\linewidth]{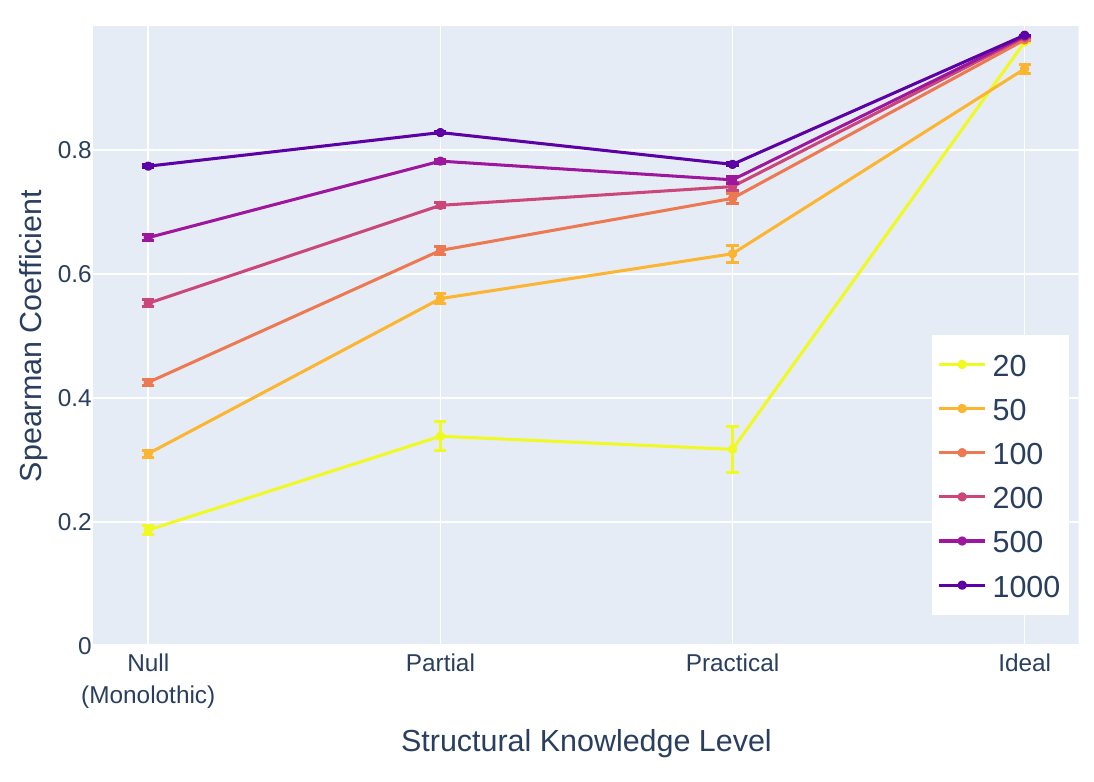}
		\caption{Hotel Reservation}
		\label{fig: Spearman Coefficnet of All For Hotel Reservation}
	\end{subfigure}
	
	\caption{
		This figure presents the average Spearman correlation coefficient, a metric for prediction accuracy, along with standard error bars. The plot illustrates the effectiveness of performance modeling for the Self-care Mobile App and the Hotel Reservation application across various levels of structural knowledge and training data sizes. The x-axis represents the four distinct Structural Knowledge Levels: Null (a black-box, monolithic approach without any structural knowledge), Partial (incorporating the logical boundary of each module), Practical (utilizing the execution graph of the application in addition to the logical boundary of each module), and Ideal (an idealized scenario with perfect structural knowledge). The colored lines correspond to a series of training data sizes, specifically the number of randomly sampled configurations measured, from a set of $\{20, 50, 100, 200, 500, 1000\}$.
	}
	\label{fig: Spearman Coefficnet of All}
\end{figure}

The results for the Self-care app, presented in Figure~\ref{fig: Spearman Coefficnet of All For Self-care}, show that leveraging structural knowledge in the Partial and Practical scenarios \textbf{did not enhance} modeling efficacy compared to the Null (black-box) approach, regardless of the training data size.

\subsection{A Second Experiment: Where Structural Knowledge Was Much More Effective}
\begin{figure}[h]
	\centering
	\includegraphics[width=\textwidth]{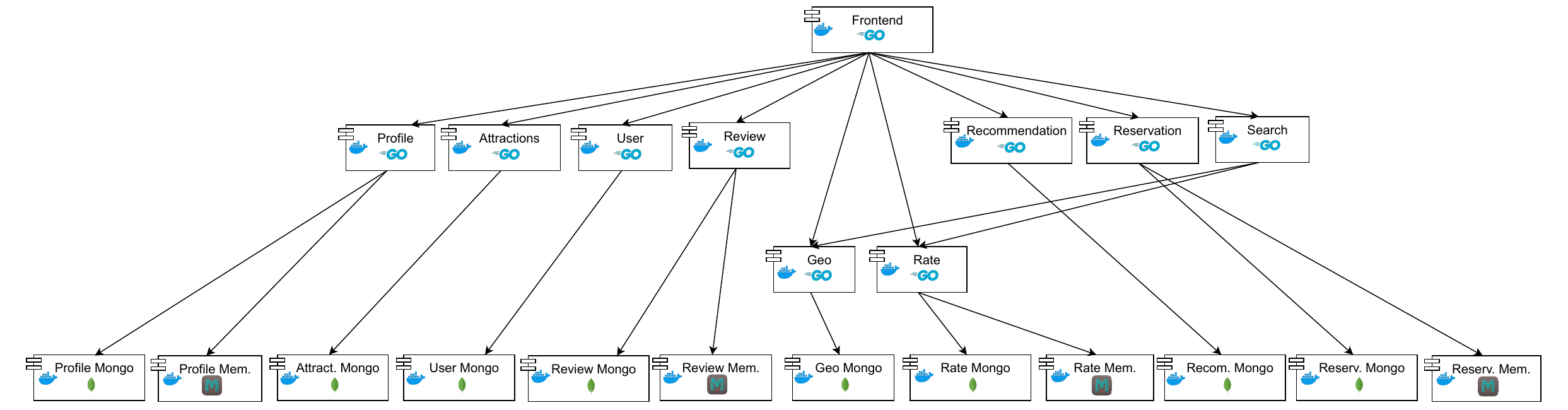}
	\caption{Execution graph of the Hotel Reservation application from DeathStarBench~\cite{gan2019open}.}
	\label{fig: hotel reservation execution graph}
\end{figure}

Following these initial findings, it was essential to determine if this outcome was a generalizable result or specific to the Self-care application. 
Therefore, the same experiment was conducted on a real-world application, the Hotel Reservation application. This application, part of the open-source DeathStarBench benchmark suite~\cite{gan2019open}, is designed to evaluate the performance of cloud-based microservices. 
Implemented in Go and leveraging gRPC, this application integrates both in-memory (Memcached) and persistent (MongoDB) back-end databases. 
It supports a hotel recommender system and reservation processing capabilities, comprising 22 interconnected modules that handle tasks such as retrieving hotel profiles, recommending hotels based on user preferences, and processing reservations. 
The system's execution graph is depicted in Figure~\ref{fig: hotel reservation execution graph}. 
The application's modules are configurable, with options varying across three distinct types: (i) Go-based microservices, (ii) Memcached as an in-memory database, and (iii) MongoDB as a persistent storage database. 
Each module type has specific configuration options with multiple possible values. 
For instance, the 10 Go-based microservices each have 3 options (e.g., \texttt{tcp\_rmem}, \texttt{tcp\_wmem}, and \texttt{cpu\_max\_ratio}), each with 6 possible values, yielding ($6^{3 \times 10} = 6^{30}$) configurations. 
The 4 Memcached services each have 3 options (e.g., \texttt{memory\_limit}, \texttt{threads}, and \texttt{slab\_growth\_factor}), also with 6 values each, contributing ($6^{3 \times 4} = 6^{12}$) configurations. 
The 8 MongoDB services each have 5 options (e.g., \texttt{tcp\_rmem}, \texttt{tcp\_wmem}, \texttt{eviction\_dirty\_target}, \texttt{eviction\_dirty\_beneath}, and \texttt{wiredTigerCacheSizeGB}), with varying numbers of values, resulting in ($6^{24} \times 5^8 \times 4^8$) configurations. 
Combined, these yield a total configuration space of approximately ($6^{66} \times 5^8 \times 4^8 \approx 2^{205}$), a figure exceeding 32 times the estimated number of atoms in the observable universe. 
The full list of options, their value ranges, and the detailed calculation are provided in Table~\ref{tab: options for hotel reservation} of Appendix~\ref{app: Option Variables}. 
These options align closely with those in prior work~\cite{liang2023modular}. 
Given its critical role in user experience, latency is a central focus of our performance modeling efforts for this system. 
As shown in Figure~\ref{fig: Spearman Coefficnet of All For Hotel Reservation}, the Hotel Reservation application demonstrated that leveraging intermediate levels of structural knowledge tangibly \textbf{improved} modeling efficacy, in contrast to the results from the Self-care application.

\subsection{What Explains The Different Experiences?}
\begin{figure}[h]
	\centering
	\includegraphics[scale=0.5]{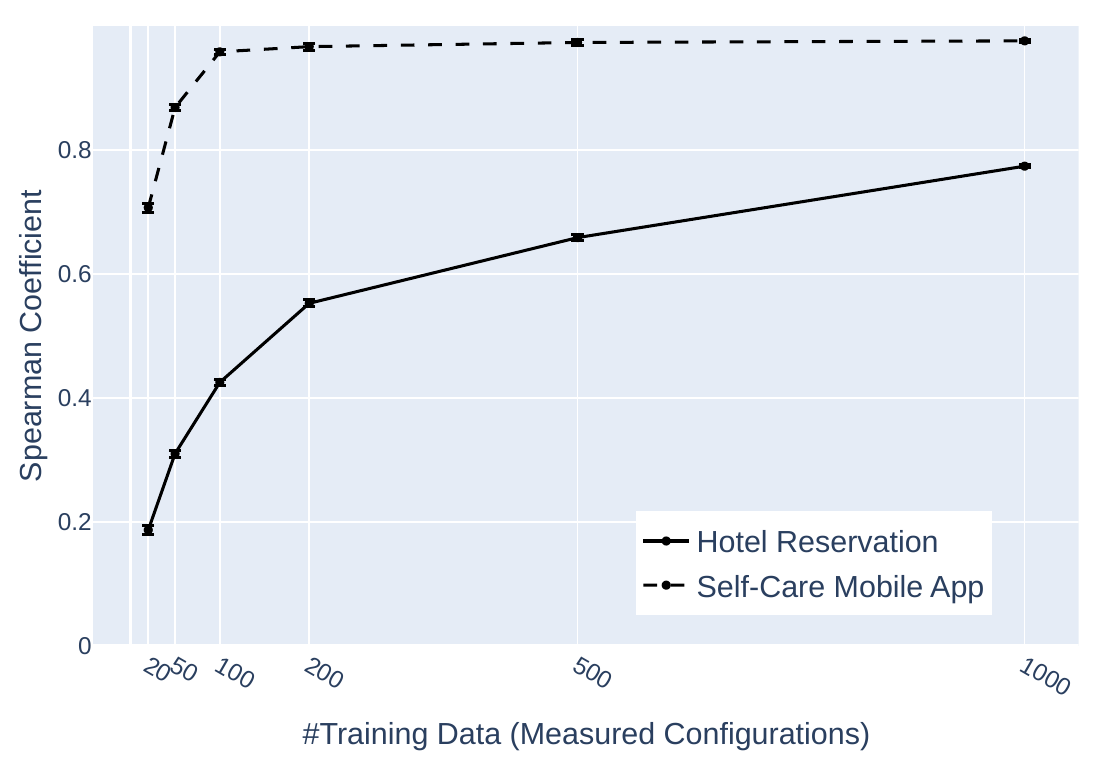}
	\caption{This figure compares the modeling hardness of the two applications by plotting the average Spearman correlation coefficient against the number of training data (measured configurations). The data presented here is the same as the ``Null'' scenario in Figure~\ref{fig: Spearman Coefficnet of All}, where a black-box performance model (specifically, random forest regression) is used without any structural knowledge.}
	\label{fig: Spearman Coefficnet of Null For Hotel Reservation}
\end{figure}

Figure~\ref{fig: Spearman Coefficnet of Null For Hotel Reservation} compares the two systems and helps us understand the significant differences between them.
This discrepancy between the two experiments raised a key question about what accounted for the difference.
A central difference between the two systems lies in their structural aspects--such as the number of modules, options per module, and inter- and intra-module interactions.
The Hotel Reservation application, with its greater complexity, benefited more from the additional structural knowledge, which is a key insight of our study.
This finding led us to the central research questions of this paper, where we explore how structural aspects affect modeling hardness and how structural knowledge and modeling hardness impact the opportunity for improvement.

\vspace{-10pt}

\section{Methodology}
\label{sec: Methodology}
This section outlines the methodological framework designed to address the research questions posed in Section~\ref{sec: intro}, which explore the influence of structural aspects and knowledge levels on the hardness and opportunity of performance modeling in modular software systems. 
Given the complexity and scale of configurable distributed systems, we employ a combination of \emph{controlled simulated experiments and statistical analysis} to systematically investigate these relationships. 
The methodology leverages \emph{synthetic system models} to isolate key variables and test hypotheses. 

\subsection{Overview of Experimental Approach}
\label{subsec: Overview of Experimental Approach}
\begin{figure}[h]
	\centering
	\includegraphics[width=\linewidth]{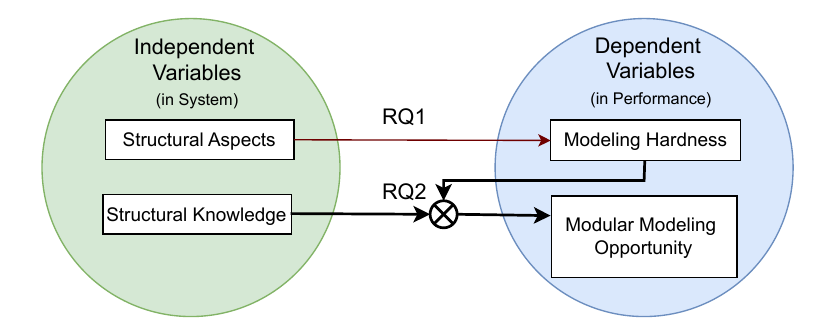}
	\caption{Relationships between independent variables, i.e., structural aspects and structural knowledge, and dependent variables, i.e., hardness and opportunities, in performance modeling, as addressed in RQ1 and RQ2.}
	\label{fig: hypotheses structure}
\end{figure}

To address the research questions outlined in Section~\ref{sec: intro}--RQ1 on how structural aspects of modular systems affect modeling hardness, and RQ2 on how structural knowledge and hardness impact the opportunity for improvement--we adopt 
a simulation approach, were we run controlled experiments with synthetic system models.
This approach enables systematic exploration of the relationships between independent variables, which are controllable during experiments, and dependent variables, which are observed in experiments. Independent variables comprises structural aspects and structural knowledge of the system. Dependent variables comprises the modeling hardness and opportunity.
Figure~\ref{fig: hypotheses structure} illustrates these relationships, highlighting the focus of RQ1 (red arrow) on the link between structural aspects and hardness, and RQ2 (bold arrows) on the interplay of structural knowledge and hardness in creating opportunities for performance modeling improvements.

The experimental framework involves generating synthetic modular software systems with varying structural properties, as defined in Section~\ref{sec: structural aspects} (e.g., the number of options per module), and simulating their performance under different levels of structural knowledge, as outlined in Section~\ref{sec: structural knowledge} (e.g., the logical boundary of each module and a superset of causal edges inside or across modules). 
By constructing performance-influence models for these systems and measuring their hardness (Section~\ref{sec: hardness}) and opportunity (Section~\ref{sec: opportunity}), we can isolate the effects of structural complexity and knowledge granularity. 

This methodology employs statistical analysis (detailed in Sections~\ref{subsec: Addressing RQ1: Structural Aspects vs. Hardness} and \ref{subsec: Addressing RQ2: Structural Knowledge and Hardness vs. Opportunity}) to quantify these relationships (represented in Figure~\ref{fig: hypotheses structure}) and evaluate their significance in addressing our research questions.
The specific design and implementation of these experiments, including dataset generation and evaluation protocols, are described in Section~\ref{sec: Experimental Setup}.

\subsection{General Structure of Configurable Modular Software Systems}
\label{subsec: Structure of Configurable Modular Software Systems}
Before delving into the details of the methodology, let's first clarify how we conceptualize a modular software system in this study.
Given that the system comprises multiple sets of variables that can influence one another in a causal manner, we adopt the terminology of causality here.
Through the lens of causality~\cite{liang2023modular, iqbal2022unicorn}, every configurable modular software system contains a set of \emph{performance variables} and multiple modules that contain multiple variables: 
\begin{itemize}
	\item[-] \emph{Option Variables.} The set of variables which are configurable by the designer or the stakeholder of the system. For example, \texttt{maxconns\_fast} option for queuing extra connections than the maximum in Memcached.
	\item[-] \emph{Intermediate Variables.} The set of variables which are affected by the value of some option variables that might be from other modules, and are not configurable. 
	For example, the memory usage of the module as an intermediate variable which can be affected by the enabling or disabling of \texttt{maxconns\_fast} option in Memcached.
	Note that the impact of multiple options on the same intermediate variables represents part of their interactions in the system. 
\end{itemize}

\noindent
The set of \emph{performance variables}, e.g., latency and energy consumption,  are defined for the entire system and are directly influenced by the subset of intermediate variables of all modules. 
Note that, some intermediate variables may affect others and indirectly influence the performance variables. 
For example, the cache-miss rate affects CPU usage, which, in turn, impacts the latency.

\begin{figure}
	\centering
	\includegraphics[width=\linewidth]{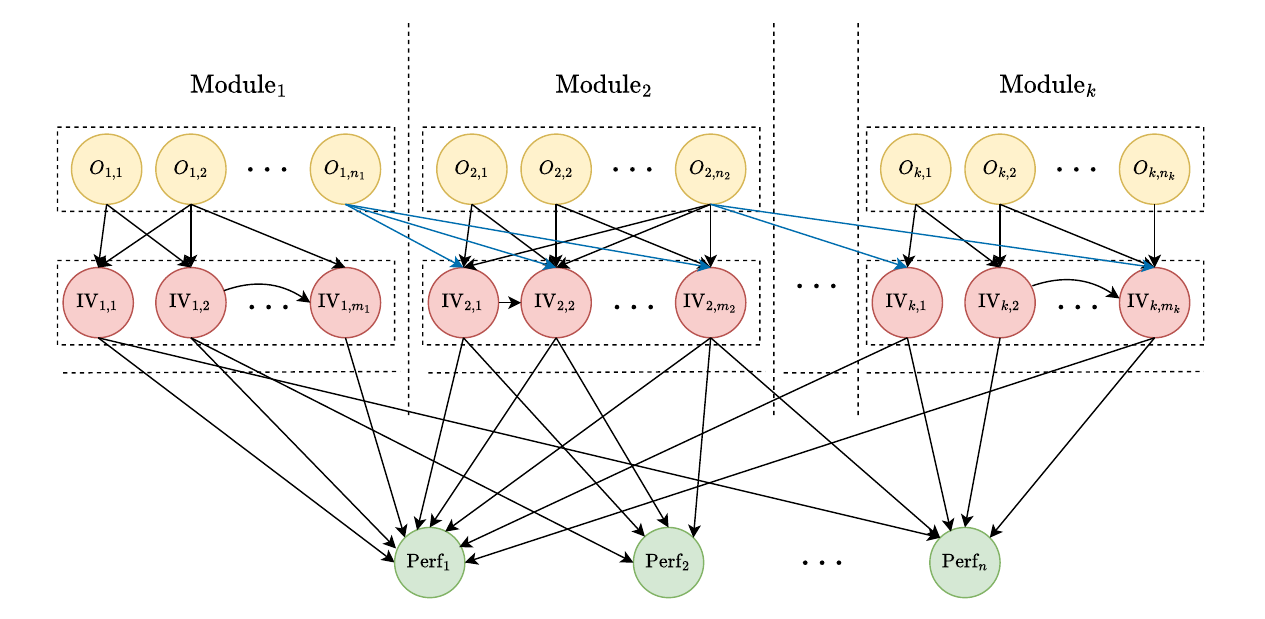}
	\caption{Example of a causal influence graph, which is directed acyclic, with all possible interactions, i.e., inter-modular (blue arrows) and intra-modular (black arrows), for the assumed configurable modular software system.
		Each circle represents a node in this graph. 
		The node of $O_{i,j}$ represents the $j$-th option variable of the $i$-th module. 
		$\mathrm{IV}_{i,j}$ represents the $j$-th intermediate variable of the $i$-th module. 
		$\mathrm{Perf}_k$ represents the $k$-th performance variable of the corresponding modular system. 
		Dashed lines indicate the logical boundary of each module.
	}
	\label{fig: general modular system model}
\end{figure}

Figure~\ref{fig: general modular system model} presents a causal influence graph illustrating the influence relationships between system variables.  
This influence structure is depicted as a directed acyclic\footnote{Acyclicity ensures that causality flows in a single direction, making it easier to interpret cause-and-effect relationships. The author of~\cite{pearl2016causal} explains how acyclic graphs help us avoid the complexity of simultaneous equations and feedback loops, simplifying causal interpretation.} graph (similar to structural causal models~\cite{kaddour2022causal}).
In the first level of the graph (i.e., top-down looking), we can see the option variables that are set by the designer/stakeholder. 
They are the root cause of changes in intermediate variables besides the influence of them on each other. 
In this figure, option and intermediate variables for each module, $\text{Module}_i$, are represented as $O_{i,j}$ and $\text{IV}_{i,j}$, respectively, while performance variables are denoted as $\text{Perf}_i$. 
The model assumes transitive causal influence; for instance, if $O_{i,j}$ influences $\text{IV}_{i,j}$ and $\text{IV}_{i,j}$ influences $\text{IV}_{l,m}$, then $O_{i,j}$ also indirectly influences $\text{IV}_{l,m}$. 

\subsection{Problem Formulation}
\label{sec: problem formuation}

To investigate the challenges and potential improvements in performance modeling of modular distributed software systems, as motivated in Section~\ref{sec: problem of hardness and opportunity}, it is essential to formally define independent and dependent variables that shape this process. 

\subsubsection{Independent Variables}
\label{sec: independent variables}
In this part, we elaborate on the formulation of independent variables, which are controlled in our designed experiments. 
\paragraph{Structural Aspect}
\label{sec: structural aspects}
In this study, we define the structural aspects of a configurable modular software system as a set of measurable properties derived from its causal influence graph (Figure~\ref{fig: general modular system model}), which represents the relationships between configuration options, intermediate variables, and performance metrics, as introduced in Section~\ref{subsec: Structure of Configurable Modular Software Systems}. These aspects--namely, the number of configuration option variables (Option\#), the rate of influence edges within a single module (\iewithinp), the rate of influence edges across modules (\ieacrossp), and the number of modules (Module\#)--are selected because they collectively characterize the structural complexity and connectivity of the system’s variables within and across its modular components. 
By quantifying these properties, we establish a systematic basis for analyzing how the architecture of a modular system impacts the hardness of performance modeling, as mentioned in RQ1 (Section~\ref{sec: intro}). The following list details these aspects:

\begin{itemize} 
	\item[-] \textbf{Option\#}. The number of configuration option variables \emph{within each module} (denoted by $n_1, n_2, \ldots, n_k$ in Figure~\ref{fig: general modular system model})). 
	For simplicity in the experiments, we set this number to be the same for all modules. For example, when we say, ``Option\# is $n$,'' it implies that this number is equal for all modules, i.e., $n = n_1 = n_2 = \cdots = n_k$.
	\item[-] \textbf{IEWithin\_p}. The probability of a single influence edge existing between any given option variable and any given intermediate variable within a module (as represented by the black arrows inside each module, in Figure~\ref{fig: general modular system model}).
	\item[-] \textbf{IEAcross\_p}. The probability of a single influence edge existing between any given option variable and any given intermediate variable across a pair of modules (as represented by the blue arrows in Figure~\ref{fig: general modular system model}). 
	\item[-] \textbf{Module\#}. The number of modules involved in the system (denoted by $k$ in Figure~\ref{fig: general modular system model})).
\end{itemize}

\noindent
For simplicity, this quantification excludes the number of intermediate and performance variables, as these are assumed to remain relatively constant. This assumption is justified by the fact that intermediate variables are typically correlated with known metrics such as CPU and memory usage, which do not vary significantly across different software variants.

The number of performance variables also remains consistent due to the complexity associated with decision-making processes involving more than three performance variables. In most cases, performance variables can be reduced to a single variable through techniques such as weighted sums or utility function definitions~\cite{deb2011multi}.

Note that the expected total number of influence edges within and across modules are calculated by multiplying the number of options (Option\#), the number of modules (Module\#), the number of intermediate variables, and the probability of a connection (IEWithin\_p and IEAcross\_p).
For example, if there are 10 options and 3 intermediate variables for 6 modules, and the probability IEWithin\_p is 0.4, the expected number of influence edges in each module is $10\times3\times0.4=12$, which is $12\times6=72$ in total.
\paragraph{Knowledge}
\label{sec: structural knowledge}

In this study, we define two distinct types of knowledge--\emph{structural knowledge} and \emph{bounding knowledge}--that are critical for analyzing the impact of system understanding on performance modeling in modular software systems, as framed by RQ2 (Section~\ref{sec: intro}). 
Structural knowledge captures the varying degrees of insight into a system’s causal influence graph (Section~\ref{subsec: Structure of Configurable Modular Software Systems}), reflecting realistic scenarios where designers may have partial or detailed awareness of module boundaries and variable interactions. 
Bounding knowledge, in contrast, establishes theoretical lower and upper limits--termed \emph{Null} and \emph{Ideal}--on the efficacy of performance-influence models, providing a baseline and ceiling against which the benefits of structural knowledge can be measured (as utilized in the analysis of the Self-care Mobile App and the Hotel Reservation applications in Section~\ref{sec: problem of hardness and opportunity}). 
Together, these knowledge types enable a comprehensive evaluation of how system insight influences modeling hardness and opportunity.

\emph{Structural Knowledge} consists of a subset of the following three elements:

\begin{itemize}
	\item \textbf{Logical Boundaries (LB)}: The set of options and intermediate variables associated with each module, indicated by dashed lines in Figure~\ref{fig: general modular system model}.
	\item \textbf{Influence Edges (IE)}: The set of all existing influence edges within and across modules, represented by arrows in Figure~\ref{fig: general modular system model}.
	\item \textbf{Potential Influence Edges (PIE)}: A superset of all possible influence edges within and across modules. This superset might be derived from domain knowledge or the system's execution graph (see, for example,~\cite{liang2023modular}). However, the exact subset of edges that actually exist in the influence graph may be unknown.
\end{itemize}

\begin{figure}
	\centering
	\includegraphics[scale=0.8]{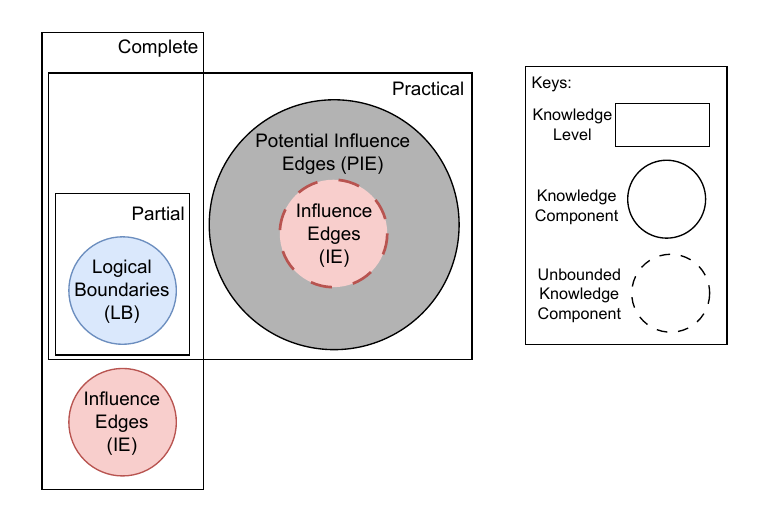}
	\caption{Relationship between different levels of structural knowledge and the corresponding knowledge elements.
		LB is the set of options and intermediate variables associated with each module, indicated by dashed lines in Figure~\ref{fig: general modular system model}.
		IE is the set of all existing influence edges within and across modules, represented by arrows in Figure~\ref{fig: general modular system model}.
		PIE is a superset of all possible influence edges within and across modules. This superset might be derived from domain knowledge or the system's execution graph. However, the exact subset of edges that actually exist in the influence graph may be unknown.
	}
	\label{fig: knowledge levels}
\end{figure}

Based on these elements, we classify structural knowledge into three levels (see Figure~\ref{fig: knowledge levels}):
\begin{itemize}
	\item[-] \emph{Partial}: Knowledge of the logical boundaries of the modules (i.e., LB) only.
	\item[-] \emph{Complete}: Full knowledge of the system's structure, including the logical boundaries and all existing influence edges (i.e., LB and IE).
	\item[-] \emph{Practical}: An intermediate level of knowledge, broader than \emph{Partial} but narrower than \emph{Complete}. This includes knowledge of the logical boundaries (i.e., LB) and a set of potential influence edges (i.e., PIE), which form a superset of the actual influence edges in the system.
\end{itemize}

\emph{Bounding knowledge} refers to two knowledge sets that are independent of the practical levels of structural knowledge. These sets serve as lower and upper bounds for the performance-influence model's efficacy under any given level of structural knowledge. Hence, they are termed \emph{bounding knowledge}. 
The two sets are:

\begin{itemize}
	\item[-] \emph{Null} knowledge: This represents a scenario where structural knowledge is not available, not even the logical boundaries of modules (the minimum level of structural knowledge). In this case, the system is treated as monolithic. The performance-influence model's efficacy under this assumption provides a likely lower bound for its performance under any level of structural knowledge.
	\item[-] \emph{Ideal} knowledge: This represents a scenario where the real values of intermediate variables for any test configuration at runtime are known without requiring measurement. Although practically unattainable, this assumption serves as an upper bound for the performance model’s efficacy under any structural knowledge level.
\end{itemize}

\subsubsection{Dependent Variables}
In this part, we elaborate on the formulation of dependent variables, which are influenced by the independent variables, and are observed and measured in our designed experiments. 
\paragraph{Hardness}
\label{sec: hardness}
In Section~\ref{sec: intro}, we introduced the concept of the \emph{hardness} of modeling a performance-influence, defined as the minimum achievable average loss relative to the amount of available training data. A natural question arises: how can this concept be quantified in practice?

To address this, we consider the effect of the order of magnitude of the training data size. 
A higher number of training data points leads to increased measurement costs. 
Monitoring the trade-off between potential improvements in model efficacy and the associated measurement cost seems essential here. 
This range of training data sizes is a common practice in the literature, and the required data is typically linearly proportional to the logarithmic factor of the system's configuration space.
Specifically, we can restrict the number of training data to several discrete levels and measure the corresponding average loss of a given learning method at each level. 
Formally, let there be $T$ ordered levels of training data sizes, denoted as $n_1 < n_2 < \cdots < n_{T-1} < n_T$. (For example, we used a range of training data sizes, from 20 to 1000, i.e., $n_1=20, n_2=50, n3=100, n_4=200, n_5=500, n_6=1000$, in Section~\ref{sec: problem of hardness and opportunity}.)

Let $l_i$ represent the minimum average loss, with respect to a loss function $L$, for a model trained on $n_i$ data points. We define the hardness of the model $\mathcal{M}$ with respect to the loss function $L$ as:

\begin{equation}
	\hard^{(L)}(\mathcal{M}) = \chard \sum_{i=1}^T \frac{l_i}{n_i}
\end{equation}

\noindent
where $\chard$  is a scaling constant that depends on the values of $n_i$s and the range of $l_i$s. 
For instance, if $L$ is bounded between 0 and 1, the hardness is at most $\chard \sum_{i=1}^T \frac{1}{n_i}$. 
To ensure that the hardness measure remains within a meaningful scale like in the range of 0 to 1, we set $\chard$ as the inverse of this summation, i.e., $\chard = 1/{\sum_{i=1}^T \frac{1}{n_i}}$.

\note{From a pragmatic perspective, computing the minimum average loss requires employing existing optimization methods, e.g., Bayesian optimizers, to search over hyper-parameters of the associated learning method. 
	Additionally, to mitigate overfitting, established techniques like 
	k-fold cross-validation, with different splits of training and test data, should be used.
	To ensure a fair comparison, all learners should receive a comparable hyperparameter search budget and be evaluated using the same cross-validation procedure.
}

This formulation provides a practical and consistent method for quantifying the hardness of a performance-influence model based on the training data size, the associated minimum average loss, and the techniques necessary to achieve reliable optimization.

\paragraph{Opportunity}
\label{sec: opportunity}
Quantifies the potential for improving a performance model's efficacy by using structural knowledge. 
It measures the ``gap'' in efficacy between a model that uses no structural knowledge (Null) and a theoretical ideal model (Ideal) that has perfect knowledge.

To formally define this, we first establish the gap ($\mathcal{G}_i^{(\mathcal{P})}(\mathcal{M})$) as the difference in a model's efficacy ($\mathcal{P}$) between the Ideal knowledge scenario ($p_i^*$) and the Null knowledge scenario ($p_i^\perp$):

\begin{equation}
	\mathcal{G}_i^{(\mathcal{P})}(\mathcal{M}) = p_i^* - p_i^\perp
\end{equation}

The Opportunity itself is then calculated by measuring how much of this gap is ``filled'' by using a specific level of structural knowledge ($\mathcal{K}$).
This is expressed in a formula that sums up the filling ratio ($f_i$) across different training data sizes ($n_i$), weighted by a scaling constant ($\copp$) for normalization.

\begin{equation}
	\opp^{(\mathcal{P}, \mathcal{K})}(\mathcal{M}) = \copp \sum_{i=1}^T \frac{f_i \times \mathcal{G}_i^{(\mathcal{P}, \mathcal{K})}(\mathcal{M})}{n_i} 
	\label{eq: Opp}
\end{equation}

In simpler terms, Opportunity measures the effectiveness of a particular level of structural knowledge ($\mathcal{K}$) by comparing its performance to a baseline (Null knowledge) and a perfect upper limit (Ideal knowledge). 
A higher opportunity value indicates that the structural knowledge provided a more significant improvement in model efficacy.

\subsubsection{Metrics}
\label{subsec: Metrics}
In this paper, for measuring the efficacy ($\mathcal{P}$) of a performance-modeling, we used two effective metrics: (i) Mean Arctangent Percentage Error (MAAPE)\cite{kim2016new} to assess accuracy, and (ii) Spearman's correlation\cite{wissler1905spearman} to evaluate ranking consistency.
MAAPE provides robustness against extreme values, addressing common issues in performance measurement, while Spearman’s correlation quantifies the monotonic relationship between predicted and actual values, ensuring that ranking order is preserved.

MAAPE is particularly useful in scenarios where absolute accuracy is critical, such as when performance predictions are used to determine whether a system meets a strict goal, e.g., the performance is greater than a given threshold. 
In contrast, Spearman’s correlation is beneficial when the goal is to preserve the ranking order of configurations, such as when selecting the top-performing system (i.e., an optimization task) configurations without requiring high numerical accuracy.

For a given set of $n$ data points with predicted values $\hat{y}_i$ and actual values $y_i$, MAAPE is defined as:

$$
\text{MAAPE} (\hat{y}_1, \ldots, \hat{y}_n; y_1, \ldots, y_n)= \frac{1}{n} \sum_{i=1}^{n} \arctan{\left(\left|\frac{y_i - \hat{y}_i}{y_i}\right|\right)}
$$

To derive an accuracy metric, we use a scaled version of MAAPE, i.e., a mapping that linearly transforms the range of MAAPE from $[0, \frac{\pi}{2}]$ to $[0,1]$:

$$
\acc(\hat{y}_1, \ldots, \hat{y}_n; y_1, \ldots, y_n) = 1 - \overbrace{\frac{2}{\pi} \text{MAAPE} (\hat{y}_1, \ldots, \hat{y}_n; y_1, \ldots, y_n)}^\text{Scaled MAAPE}
\quad .$$

Spearman's correlation coefficient is computed as:

$$
\text{Spearman Coefficient}(\hat{r}_1, \ldots, \hat{r}_n; r_1, \ldots, r_n) = \frac{\Sigma_i(\hat{r}_i - \hat{\mu})(r_i - \mu)}{\sqrt{(\Sigma_i(\hat{r}_i-\hat{\mu})^2)(\Sigma_i(r_i-\mu)^2)}}
$$

\noindent
where $\hat{r}_i$ and $r_i$ denote the ranks of predicted and actual values, respectively, and $\hat{\mu}$ and $\mu$ are their mean ranks.

Accordingly, we define two loss functions, particularly used in the measuring of the hardness (Section~\ref{sec: hardness}), based on these two efficacy metrics that are ``$1-\acc$'' (that is equal to the scaled MAAPE) and ``$1 - \text{Spearman Coefficient}$''. (Note that all observed coefficients are positive.)


\subsection{Addressing RQ1: Structural Aspects vs. Hardness}
\label{subsec: Addressing RQ1: Structural Aspects vs. Hardness}
The objective of RQ1 is to investigate the impact of each structural aspect on the hardness metric. 
It seems that a plausible approach to address RQ1 is to frame it as a (polynomial) \emph{regression problem}, where the structural aspects serve as independent variables. 
This approach enables the quantification of each aspect's contribution to hardness, as determined by their coefficients in the regression model.

To construct such a regression model, it is necessary to gather data on hardness values corresponding to various combinations of structural aspect variables. 
Each combination can be represented as a quadruple of the form $\langle\text{Option\#, \iewithinp, \ieacrossp, Module\#}\rangle$. 
The dataset corresponding to each combination consists of sampled configurations along with their respective intermediate and performance variables.
By measuring the hardness based on this dataset, we obtain a set of tuples, each consisting of a combination and its corresponding hardness value.

From a statistical analysis perspective, the variability in dataset collection is essential to ensure that the regression model accurately captures the relationships between structural aspects and hardness.
How can such a dataset be collected practically to ensure sufficient variability with respect to the structural aspects? Addressing this question is part of the focus of the experimental setup in Section~\ref{sec: Experimental Setup}.

\subsection{Addressing RQ2: Structural Knowledge and Hardness vs. Opportunity}
\label{subsec: Addressing RQ2: Structural Knowledge and Hardness vs. Opportunity}
\begin{figure}
	\centering
	\includegraphics[scale=0.5]{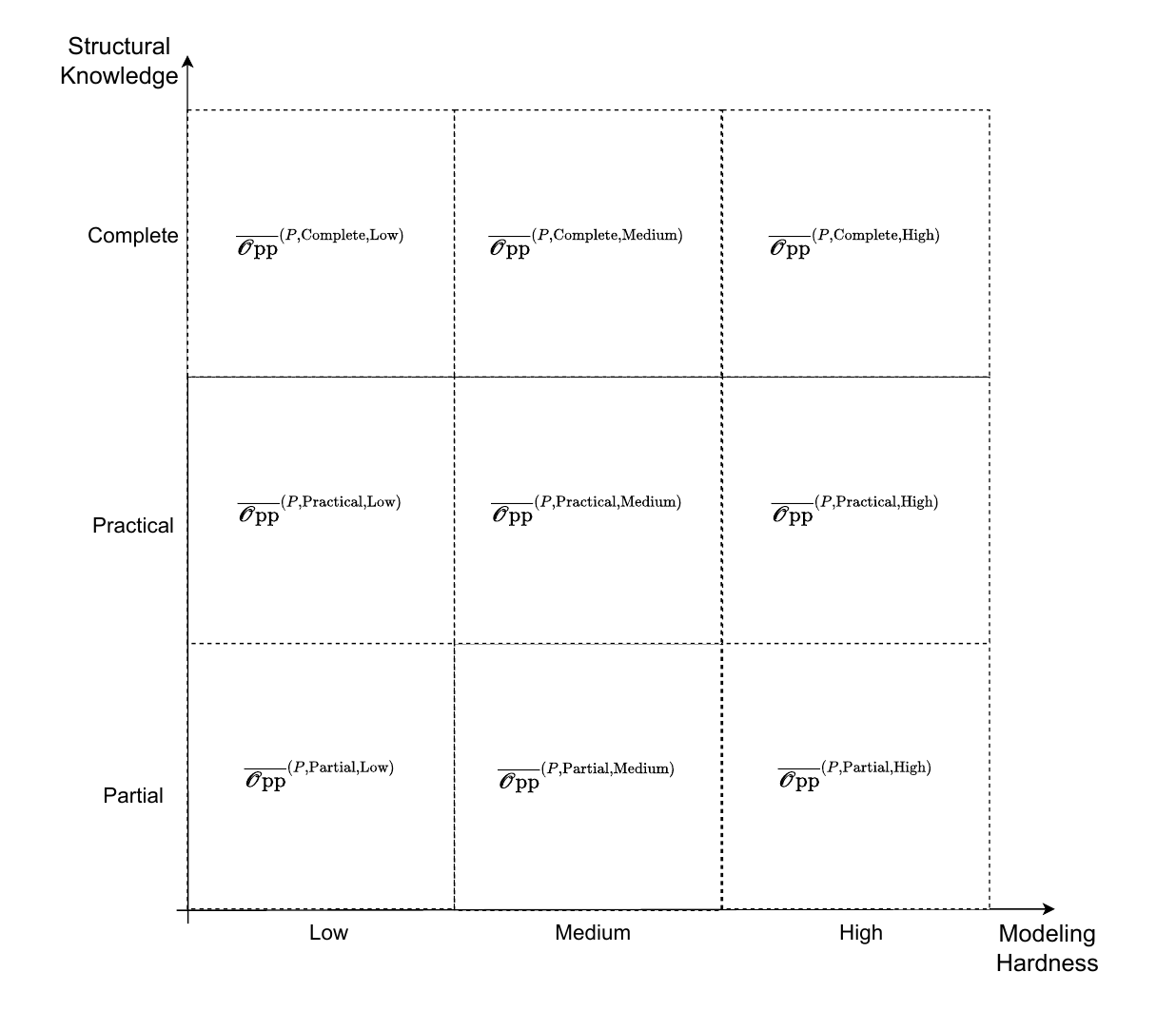}
	\caption{This analytical matrix is a framework used to formulate and investigate RQ2. 
		It systematically organizes the key variables to explore the relationship between the levels of structural knowledge (Partial, Practical, and Complete) and the modeling hardness level (Low, Medium, and High) of an application.
		Each cell in the matrix represents the average opportunity derived from the dataset detailed in Section~\ref{sec: Experimental Setup} (based on Equation~\ref{eq: Opp}).
		For instance, $\avgopp^{(\mathcal{P},\text{Complete},\text{Low})}$ denotes the average measured opportunities across all sampled system setups where the modeling hardness, in terms of $\mathcal{P}$, is Low, given a Complete level of structural knowledge.
	}
	\label{fig: raw analytical matrix}
\end{figure}

To address RQ2, we introduce an analytical matrix, as illustrated in Figure~\ref{fig: raw analytical matrix}. This approach is motivated by the need to analyze the complex interplay between modeling hardness (a continuous metric) and structural knowledge (a categorical variable). 
While it is mathematically possible to model the relationship between these variables continuously using techniques such as ANalysis of COVariance (ANCOVA), this approach would not align with our primary goal of producing actionable and interpretable insights for practitioners.
Unlike modeling hardness, which is an inherent, system-specific property that can be represented by a single metric $\mathcal{P}$, opportunity is a more complex, conditional value. 
It represents the potential for improvement, contingent on both the specific level of structural knowledge available and the system's inherent hardness. 
A continuous model would produce a numerical prediction for opportunity, requiring practitioners to then define arbitrary thresholds to make that value useful. 
This would reintroduce subjectivity and undermine the direct utility of the model.

In contrast, our analytical matrix is designed from the outset to produce decisionable categories. 
This is a well-established and practical approach in software engineering research, as seen in studies~\cite{shatnawi2010quantitative} and  \cite{arcelli2016comparing}. 
The primary motivations for this quantization are to:
\begin{itemize}
	\item \emph{Enhance Interpretability and Actionability}: Converting continuous metrics into categorical ratings (e.g., ``low,'' ``medium,'' and ``high'') makes the results more accessible and easier to communicate to non-statistical stakeholders, such as managers and engineering teams~\cite{heitlager2007practical}. 
	This enables the creation of actionable, interpretable rules for quality control and risk mitigation~\cite{shatnawi2010quantitative}. 
	\item \emph{Support Structured Hypothesis Testing and Data Visualization}: The matrix provides a structured way to formulate and test our hypotheses by comparing the average ``opportunity'' values in different cells. 
	This structure is also useful basis for a visual representation, which can effectively summarize the complex relationships between the variables and make the findings accessible to a broader audience.
	\item \emph{Operationalize Findings}: Many practical outcomes in software engineering are treated as binary (e.g., fault-prone vs. not). Using cutpoints to create categorical variables allows for the development of practitioner-friendly, explainable classification rules without relying on black-box models~\cite{arar2016deriving}.
\end{itemize}

For this study, we specifically partition the finite range of our hardness metric, into four equal intervals: the first quartile represents Low hardness, the second and third quartiles constitute Medium hardness, and the final quartile corresponds to High hardness. 
This allows for a robust and interpretable analysis that is compatible with the categorical levels of structural knowledge.

Each cell in the matrix represents the average opportunity derived from the dataset detailed in Section~\ref{sec: Experimental Setup} (based on Equation~\ref{eq: Opp}).
Each data tuple, characterized by a combination of modeling hardness and structural knowledge level, maps uniquely to one of the matrix cells. 
For instance, $\avgopp^{(\mathcal{P},\text{Complete},\text{Low})}$ denotes the average measured opportunities across all sampled system setups where the modeling hardness, in terms of $\mathcal{P}$, is Low, given a Complete level of structural knowledge.

RQ2 is subsequently formulated as a set of statistical hypotheses based on comparisons of the average opportunities across matrix cells. These hypotheses are derived directly from the preliminary observations and research questions introduced in Section~\ref{sec: problem of hardness and opportunity}. 
The initial discrepancy observed between the ``Self-care Mobile App'' (lower hardness, lower opportunity) and the ``Hotel Reservation'' application (higher hardness, higher opportunity) motivated the central research questions of this study. 
The hypotheses below provide a formal and testable framework for exploring the relationship between these variables and generalizing the findings beyond the initial case studies. 
Given that the matrix comprises nine cells, 36 pairwise null hypotheses can be defined, i.e., $C(9,2)=\frac{9\times 8}{2}=36$. Each null hypothesis has a corresponding alternative hypothesis, formulated according to the following expectations:

\begin{itemize}
	\item \textbf{Increased modeling hardness provides greater opportunities for leveraging modular knowledge (at any level of structural knowledge).}
	This expectation pertains to row-wise comparisons. For instance, for the null hypothesis 
	$$H_0: \avgopp^{(\mathcal{P},\text{Complete},\text{Low})} = \avgopp^{(\mathcal{P},\text{Complete},\text{Medium})} \quad,$$ 
	the alternative hypothesis is formulated as
	$$H_a: \avgopp^{(\mathcal{P},\text{Complete},\text{Low})} < \avgopp^{(\mathcal{P},\text{Complete},\text{Medium})} \quad.$$
	\item \textbf{Higher levels of structural knowledge yield greater opportunities.}
	This expectation relates to column-wise comparisons. For example, for the null hypothesis
	$$H_0: \avgopp^{(\mathcal{P},\text{Medium},\text{Low})} = \avgopp^{(\mathcal{P},\text{Complete},\text{Low})} \quad,$$ 
	the alternative hypothesis is expressed as
	$$H_a: \avgopp^{(\mathcal{P},\text{Medium},\text{Low})} < \avgopp^{(\mathcal{P},\text{Complete},\text{Low})} \quad.$$
	\item \textbf{The influence of modeling hardness on opportunity is not equivalent to the influence of structural knowledge.}
	This assumption extends to comparisons that go beyond simple row-wise or column-wise analyses. For example, consider the null hypothesis:
	$$H_0: \avgopp^{(\mathcal{P},\text{Complete},\text{Medium})} = \avgopp^{(\mathcal{P},\text{Partial},\text{High})} \quad,$$ 
	The corresponding alternative hypothesis is:
	$$H_a: \avgopp^{(\mathcal{P},\text{Complete},\text{Medium})} \neq \avgopp^{(\mathcal{P},\text{Partial},\text{High})} \quad.$$
	If the null hypothesis is rejected, one approach to identify the more influential factor is to quantify the effect size for each group using established metrics, e.g., Common Language Effect Size (CLES).
\end{itemize}


\subsection{Experimental Setup}
\label{sec: Experimental Setup}
In this section, we explain the details of making the synthetic dataset, learning methods used for performance modeling of each level of structural knowledge, and the statistical methods used to address RQ1 and RQ2. 

\subsubsection{Synthtetic Dataset Creation}
\label{subsec: Synthtetic Dataset Creation}
This section describes the construction of a synthetic dataset to address the research questions. 
The synthetic dataset serves as the foundation for the controlled experiments proposed in Section~\ref{subsec: Overview of Experimental Approach}, enabling the regression analysis of structural aspects’ impact on hardness (Section~\ref{subsec: Addressing RQ1: Structural Aspects vs. Hardness}) and the matrix-based evaluation of structural knowledge and hardness on opportunity (Section~\ref{subsec: Addressing RQ2: Structural Knowledge and Hardness vs. Opportunity}). By generating modular system models with controlled structural properties and knowledge levels, we create a reproducible environment to test the hypotheses underlying our research questions.

To construct this synthetic dataset for robust statistical analysis, we define a parameter space for structural aspects, as outlined in Section 4.3.1, by reviewing microservice-based systems in existing literature, including the DeathStarBench benchmark suite (comprises Hotel Reservation, Social Network, and Media Service application)~\citep{gan2019open}, Google Online Boutique application~\cite{googleBoutique} with 11 microservices, and other studies on modular distributed systems~\cite{quin2022seabyte, liang2023modular}. 
These works describe systems with module counts ranging from 10 to 41 and configuration options per module typically between 4 and 15, reflecting common architectural patterns in cloud-based microservices. The dataset’s ranges---5 to 40 modules (Module\#) and 6 to 16 binary configuration options per module (Option\#)---cover over most of the surveyed systems, slightly broadening the lower bound to include smaller systems while excluding outliers like Netflix’s architecture with thousands of modules~\cite{bogner2019microservices}. The construction process involves three key steps: (i) sampling the number of modules and configuration options per module, (ii) probabilistically generating influence edges within (\iewithinp) and across (\ieacrossp) modules to form a causal influence graph, and (iii) simulating performance data by sampling configurations and computing intermediate and performance variables using random polynomial functions. These steps ensure sufficient variability to capture relationships between independent and dependent variables, while maintaining computational feasibility for controlled experiments.

\textit{Number of Options} (Option\#). 
To simplify the representation without loss of generality, we assume that all configuration options are binary. This assumption is justified because any non-binary option in real-world applications can typically be decomposed into a set of binary options representing its valid quantized values~\cite{gong2022does}. For instance, consider a microservice configuration option \texttt{threads} with a range of valid values among the set of $\{2,5,8,11\}$. 
This can be equivalently expressed as four binary options: \texttt{threads\_2}, \texttt{threads\_5}, \texttt{threads\_8}, and \texttt{threads\_11}. Setting one of these binary options to \texttt{True} indicates that \texttt{threads} is configured to the corresponding value. This binary encoding approach aligns with how configuration options are commonly represented in distributed systems, such as the Hotel Reservation application in DeathStarBench~\cite{gan2019open}, as discussed in~\cite{liang2023modular}.
Reflecting practical limits observed in distributed applications, we assume the number of binary configuration options per module ranges between 6 and 16. 

\textit{Number of Modules} (Module\#).
The number of modules per system is assumed to vary between 5 and 40. Accordingly, the largest system in the dataset includes up to 640 binary options (i.e., $40 \times 16$), resulting in a configuration space size of $2^{640}$.

\textit{Influence Edges Within Modules} (\iewithinp). Unlike Option\# and Module\#, which are assigned static values, the distribution of influence edges within modules is modeled probabilistically. 
This is because the set of connections between options and intermediate variables is not fixed in real-world applications, though its magnitude can be controlled via a corresponding parameter. 
To capture this structural aspect, we define a uniform distribution parameterized by a probability factor $p_w$, which governs the likelihood of a connection between any given pair of options and intermediate variables within a module. 
The range of $p_w$ is set to [0.5, 1.0], where $p_w = 1.0$ indicates that all options influence all intermediate variables within each module.
This range is selected to avoid structurally sparse or disconnected configurations that may arise with lower probabilities, particularly given the small number of intermediate variables (here, 3). Excessively sparse graphs in this context are considered unrealistic and may not accurately reflect practical scenarios.

\textit{Influence Edges Across Modules} (\ieacrossp).
Similar to \iewithinp, influence edges across modules are modeled using a probability distribution. 
Specifically, a truncated normal distribution, parameterized by $\mu_a$ and $\sigma_a$, both restricted to the interval $[0.01, 0.4]$, is employed to define the probability of a connection between options and intermediate variables across different modules. 
The choice of a truncated normal distribution within this range is motivated by three key factors: (i) the rate of influence edges across modules is typically lower than within modules due to modularity constraints, hence the upper bound of 0.4; (ii) the variability in influence edges across modules depends on the specific roles of the modules; and (iii) truncation ensures that probabilities remain within a realistic finite range, avoiding the unrealistic infinite range of a standard normal distribution.
Note that, we assumed all intermediate variables can influence the performance variable.

To construct a system setup, i.e., the causal influence graph,  we uniformly sample the parameter space defined above for each structural aspect at random. 
These value spaces are summarized in Table~\ref{tab: params for dataset}, which outlines the corresponding parameters for each element of the quadruple $\langle\text{Option\#, \iewithinp, \ieacrossp, Module\#}\rangle$. 
For example, a sampled setup of $\langle$Option\#: 6, $p_w$: 0.6, $\langle\mu_a$: 0.3, $\sigma_a$: 0.1$\rangle$, Module\#: 10$\rangle$ represents a valid instance drawn from the specified parameter spaces.

\begin{table}[h!]
	\begin{center}
		\caption{Ranges for structural aspects and their corresponding distribution parameters, defining the space of possible modular distributed systems in the synthetic dataset.}
		\label{tab: params for dataset}
		\begin{tabular}{| *4{>{\centering\arraybackslash}m{1in}|} @{}m{0pt}@{}}
			\hline
			\shortstack{\textbf{Structural}\\ \textbf{Aspect}} & \shortstack{\textbf{Distribution} \\ \textbf{Type}}& \shortstack{\textbf{Distribution} \\ \textbf{Parameters} }       & \textbf{Value Space} \\ \hline
			\text{Option\#}      & \cellcolor{gray}     &  \cellcolor{gray}     & $\{6..16\}$      \\ \hline
			\iewithinp      & Uniform & $p_w$            & $[0.5, 1]$      \\ \hline
			\multirow{2}{*}{\ieacrossp}  & \multirow{2}{*}{\shortstack{Truncated Normal}} & \multirow{1}{*}{$\mu_a$} & \multirow{1}{*}{[0.01, 0.4]} \\  
			&    & \multirow{1}{*}{$\sigma_a$}                              &    
			\multirow{1}{*}{[0.01, 0.4]} \\ \hline
			\text{Module\#}     & \cellcolor{gray} & \cellcolor{gray}  & $\{5..40\}$      \\ \hline
		\end{tabular}
	\end{center}
\end{table}

Once the system structure is defined, a causal influence graph is generated (e.g., Figure~\ref{fig: general modular system model}). However, this graph lacks quantitative impact functions for connected elements. While it identifies influential intermediate and option variables, the magnitude of their impact remains unspecified.

\paragraph{Intermediate And Performance Variables} 
To determine intermediate and performance variable values, we first compute intermediate variables based on their connected options or other intermediate variables. 
If an intermediate variable is solely connected to option variables, a random polynomial is generated using uniform random weights from $[0.0, 1.0]$ and incorporating first- and second-order interactions. 
To ensure experimental reproducibility, this randomization process is seeded uniquely for each trial (e.g., 40 trials for our sensitivity analysis). 
Given that the influence graph is a directed acyclic graph, computations start with independent intermediate variables and proceed iteratively to dependent ones. 
Performance metrics are then computed using a random linear formula incorporating all connected intermediate variables.
A small perturbation, represented by a uniform random variable bounded within 5\,\% of the original formula's value around zero, is added to the computation to model measuring noise.
This simulation approach aligns with findings from study~\cite{siegmund2015performance}, which suggest that interactions in configurable software systems are typically of at most degree 4, enabling accurate performance modeling via polynomial regression of degree 4.

\begin{figure}
	\centering
	\includegraphics[scale=0.7]{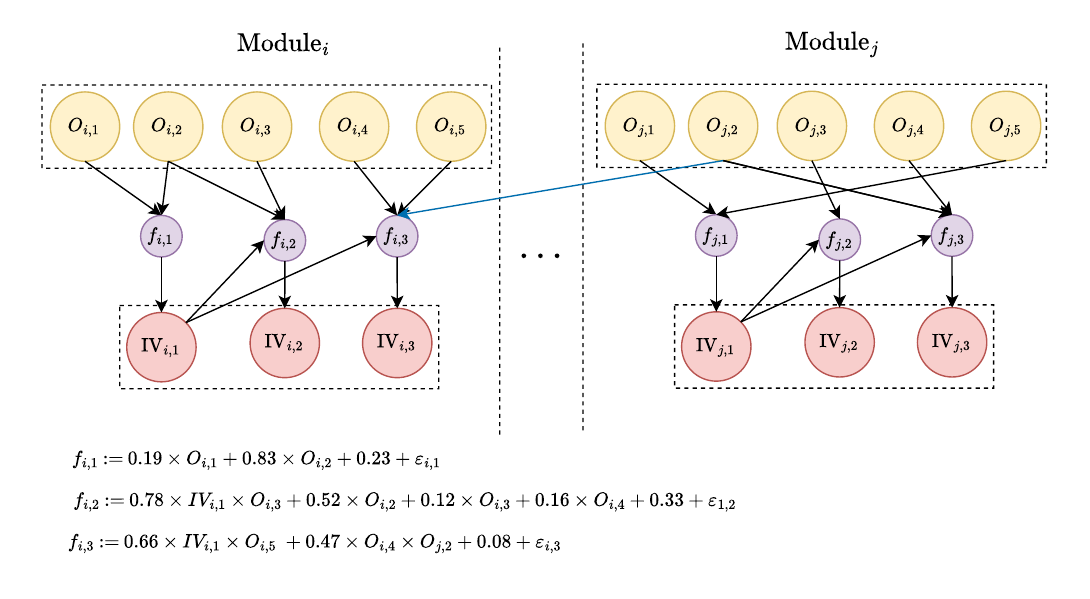}
	\caption{Example of a modular system from the synthetic dataset, showing formulas for a module’s intermediate variables. This illustrates the causal influence graph from Figure~\ref{fig: general modular system model}, where functions represent the relationships between nodes and their causal connections.   
	}
	\label{fig: example env}
\end{figure}

Finally, for each system setup, we generate the dataset by uniformly sampling the binary option configuration space at random and computing the corresponding intermediate and performance variables using the generated polynomial formulas. 
This process allows us to create datasets necessary for evaluating system hardness and opportunity.
This dataset consists of 400 samples of modular systems (e.g., a part of a sampled modular system is represented in Figure~\ref{fig: example env}), each with 2000 measured configuration samples (i.e., 1000 samples for training data and 1000 samples for testing). 

\subsubsection{Learning Methods for Performance Modeling}
\label{subsec: Learning Methods for Performance Modeling}
This part describes the learning methods employed for performance modeling across different levels of structural knowledge (Null, Partial, Practical, and Ideal, as defined in Section~\ref{sec: structural knowledge}, Figure~\ref{fig: knowledge levels}) and explains how these methods access and utilize the structural aspects of modular software systems, including the role of intermediate variables. 
The methods are applied to synthetic system models and are designed to reflect realistic scenarios, such as the performance modeling of the Hotel Reservation application (Section~\ref{sec: problem of hardness and opportunity}).

\paragraph{Accessing and Utilizing Structural Knowledge.}
The structural knowledge available to each modeling approach corresponds to the elements defined in Section~\ref{sec: independent variables}: Logical Boundaries (LB), Influence Edges (IE), and Potential Influence Edges (PIE). These elements are derived from the causal influence graph (Figure~\ref{fig: general modular system model}), which represents the relationships between configuration option variables, intermediate variables, and performance variables. 
The modeling approaches leverage this structure as follows:

\begin{itemize}
	\item[-] \emph{In The Level of Null Knowledge (Black-Box Approach):} This approach assumes no structural knowledge, treating the system as a monolithic entity. 
	It uses a single random forest regressor to model the relationship between configuration options (inputs) and performance variables (outputs), such as latency, without considering module boundaries or intermediate variables. The random forest regression is chosen for its widespread validation in prior research~\cite{grebhahn2019predicting, velez2020configcrusher, kaltenecker2020interplay} and interpretability relative to other learning methods. 
	\item[-] \emph{In The Level of Partial Knowledge:} This approach incorporates knowledge of Logical Boundaries (LB), i.e., the association of configuration options and intermediate variables with specific modules. 
	Each module is modeled individually using a random forest regressor to predict its intermediate variables (e.g., module-specific CPU or memory usage) based on its configuration options. 
	These module-level predictions are then aggregated using an additional random forest regressor to estimate system-level performance variables. 
	This hierarchical structure leverages module boundaries to reduce the complexity of the configuration space.
	\item[-] \emph{In The Level of Practical Knowledge:} This approach extends Partial knowledge by including Potential Influence Edges (PIE), a superset of causal relationships derived from the system’s execution graph (e.g., Figure~\ref{fig: hotel reservation execution graph} for the Hotel Reservation application). 
	A structural causal model~\cite{pearl2016causal, kaddour2022causal} is employed, where random forest regressors model the relationships between configuration options and intermediate variables, both within and across modules. 
	To enhance model performance, irrelevant causal edges are pruned using statistical tests, such as the Fisher-Z test, as applied in the Hotel Reservation application (Section~\ref{sec: problem of hardness and opportunity}). 
	This approach captures inter-module interactions, might improve prediction accuracy by accounting for dependencies, such as how a configuration option in one module affects intermediate variables in another.
	\item[-] \emph{In The Level of Ideal Knowledge:} This approach represents an upper bound where the true values of intermediate variables (e.g., runtime CPU and memory usage) are known for any configuration without measurement. 
	A random forest regressor uses these intermediate variables directly as inputs to predict performance variables, bypassing the need to model configuration-to-intermediate relationships. 
	This approach maximizes efficacy but is impractical due to infeasibility of obtaining the true value of all real-time intermediate variables without measurement.
\end{itemize}

\note{
	The choice of learning method for the Null and Ideal models should be the same to ensure a consistent basis for comparison between the lower and upper bounds modeling efficacy. Moreover, while other learning algorithms were possible other than the random forest, their specific effect on the obtained results is an extension of this work and is not the main focus of this paper, which is to introduce the concept of hardness and opportunity itself.
}

\paragraph{Role of Intermediate Variables.} 
Intermediate variables, such as CPU usage, memory consumption, or other module-specific metrics, are critical in modular performance modeling as they mediate the causal relationships between configuration options and system-level performance variables (Section 4.2, Figure 5). 
In the Partial and Practical knowledge scenarios, intermediate variables are explicitly modeled to capture module-specific behaviors and inter-module dependencies. 
For instance, in the Hotel Reservation application (Section~\ref{sec: problem of hardness and opportunity}), intermediate variables like memory usage in Memcached or MongoDB modules are predicted based on configuration options (e.g., \texttt{memory\_limit} or \texttt{wiredTigerCacheSizeGB}). 
These predictions inform the system-level performance model, reducing the reliance on extensive configuration sampling. 
In contrast, the Null approach ignores intermediate variables, leading to higher hardness due to the need for more training data to capture complex relationships. 
The Ideal approach assumes direct access to intermediate variable values, simplifying the modeling task but requiring unrealistic data availability. 
By incorporating intermediate variables, Partial and Practical approaches bridge the gap between Null and Ideal scenarios, leveraging structural knowledge to enhance model efficacy, as evidenced by higher efficacy in the Hotel Reservation application (Figure~\ref{fig: Spearman Coefficnet of All For Hotel Reservation}).

\subsubsection{Statistical Methods for Addressing RQ1}
\label{subsec: Addressing RQ1 Statistical Methods}
As explained in Section~\ref{subsec: Addressing RQ1: Structural Aspects vs. Hardness}, a polynomial regression was made for the collected data of all 400 sampled systems to explore the relationship between structural aspects of modular systems (explained in Section~\ref{sec: independent variables}) and their impact on the hardness of performance modeling. 
For this regression, we use the LASSO method (with an L1-norm regularization to prevent overfitting) that is optimized by a grid search on polynomial degree from 1 to 4 and regularization strength $\alpha$ in the range of 0.0001 to 10 with 500 steps (with the best mean squared error $3.14\times10^{-4}$ and $0.283$ for the regression modeling of hardness corresponding to correlation and accuracy, respectively).
For making the regression, the value of all corresponding structural aspects of each data record is scaled into the range 0 to 1 (see the importance of scaling features in machine learning methods in~\cite{sharma2022study}), based on the minimum and maximum of their value space, mentioned in the last column of Table~\ref{tab: params for dataset}.

\subsubsection{Statistical Methods for Addressing RQ2}
\label{subsec: Addressing RQ2 Statistical Methods}

To test the hypotheses related to RQ2, it is important to note that hardness is itself a dependent variable, alongside opportunity.
Therefore, we employ the most accurate regression model identified in the previous analysis in the first stage (Stage 1, represented by the red arrow in Figure~\ref{fig: hypotheses structure})--namely, the LASSO method described in Section~\ref{subsec: Addressing RQ1 Statistical Methods}.
In the second stage, we use this model to estimate the values of hardness, which are incorporated into the hypothesis testing framework (Stage 2, represented by black arrows in Figure~\ref{fig: hypotheses structure}).
This procedure, i.e., Stage 1 and 2, aligns with the well-established statistical technique known as Two-Stage Least Squares (2SLS)~\cite{angrist2009mostly}.

To ensure the validity and accuracy of the statistical tests used to address RQ2, the following assumptions are considered:

\begin{itemize}
	\item All 400 samples representing the setup configurations of modular software applications are independent, in accordance with our sampling methodology.
	\item No assumption of normality is imposed on the distribution of opportunity values across the different groups in the analytical matrix.
\end{itemize}

Given these assumptions, the Mann–Whitney U test is an appropriate non-parametric method for testing the hypotheses associated with RQ2. This test is particularly suitable when the assumption of normality does not hold, and the samples are independent.

To complement the hypothesis testing and provide a measure of practical significance, we calculate the Common Language Effect Size (CLES). This effect size expresses the probability that a randomly selected value from one group will exceed a randomly selected value from another group. Values closer to 1 indicate stronger effects.


\subsection{Threats to Validity}
\label{sec:threats-to-validity}
The study of hardness and opportunity in modular performance modeling for distributed software systems faces multiple validity threats. 
Accordingly, we adhere to the guidelines outlined in~\cite{wohlin2012experimentation, wieringa2014design}.

\subsubsection{External Validity}
External validity is the primary concern for the generalizability of our findings to real-world modular software systems. 
This study should be seen as an initial exploration of the phenomenon, rather than a comprehensive study of real-world systems. The findings are most applicable to highly configurable, loosely coupled systems like microservices and may not generalize to systems with tightly coupled modules or different architectures.

This limitation is the result of a deliberate design choice. 
Our use of a tightly controlled synthetic simulation environment allows for precise manipulation and measurement of structural variables. 
We prioritized this approach because the effects we aimed to study are subtle and require a high degree of control to isolate. 
A real-world setting would introduce confounding variables--such as dynamic workloads, hardware variability, and network dynamics--that would obscure the specific relationships between structural aspects and modeling challenges. 
Furthermore, conducting controlled experiments across the many real-world systems needed for this study would be prohibitively expensive and logistically infeasible~\cite{meiklejohn2021service}. 
By accepting this tradeoff, we were able to establish a robust foundation for understanding the fundamental dynamics of modular performance modeling, which can inform future research and application in more complex, real-world contexts.

\subsubsection{Conclusion Validity}
Conclusion validity concerns the ability to draw correct statistical conclusions about the relationships between independent variables (structural aspects and levels of structural knowledge) and dependent variables (hardness and opportunity) in our experiments. A primary threat is \emph{low statistical power}, which could arise from insufficient variability in the synthetic dataset used to model modular software systems. 
Despite our best efforts to mitigate variability issues through a comprehensive synthetic dataset (Section~\ref{subsec: Synthtetic Dataset Creation}), we cannot entirely eliminate the possibility that real-world system complexities may introduce additional variability not captured in our study.

\subsubsection{Construct Validity}
Construct validity evaluates whether the experimental setup accurately measures the intended constructs, such as hardness, opportunity, and structural knowledge. 
A significant threat is \emph{inadequate pre-operational explication of constructs}. 
The definition of hardness as the minimum average loss relative to training data size (Section~\ref{sec: hardness}) may not fully capture the practical challenges of modeling complex systems, as it depends on specific efficacy metrics. 
Similarly, categorizing structural knowledge into Partial, Practical, and Complete levels may oversimplify the spectrum of system understanding. 
Despite our best efforts to align hardness and opportunity with established definitions, and validate them against the Hotel Reservation application, we cannot entirely eliminate the possibility that alternative metric formulations may highlight different modeling challenges.

\subsubsection{Theoretical Validity}
Theoretical validity assesses alignment with the causal machine learning framework used to model performance-influence relationships (Figure~\ref{fig: general modular system model}).
A key threat is \emph{model misspecification}, where the causal influence graph (Figure~\ref{fig: general modular system model}) may not capture true relationships, such as feedback loops, due to the assumption of acyclicity. 
We adopted established causal modeling techniques (e.g., \cite{pearl2016causal}) and pruned irrelevant edges using the Fisher-Z test (Appendix~\ref{appendix: Details of evaluation results}). 
Despite our best efforts to ensure accurate causal modeling, we cannot entirely eliminate the possibility that unmodeled feedback loops affect performance relationships.

\emph{Failure to test alternative hypotheses} is another threat, as we focused on structural aspects and knowledge levels without exploring other factors like algorithm and sampling method choice.
We used robust random forest regression (i.e., a common, powerful, and capable learning model) as a primary learning model, uniform random as a common sampling method, and included robustness checks (e.g., multiple trials).
Despite our effort to focus on causal modeling, we cannot entirely eliminate the risk that alternative methods may yield different insights.

\subsubsection{Interaction Validity}
Interaction validity concerns conclusions when treatments interact with other factors, such as system complexity or knowledge granularity. The \emph{interaction of treatment and setting} is a threat, as the effect of structural knowledge (e.g., Practical vs. Partial) may vary with system complexity (e.g., number of modules). 
Practical knowledge, including the execution graph, may be more beneficial in systems with many inter-module edges (\ieacrossp). We analyzed diverse synthetic systems (Section~\ref{subsec: Synthtetic Dataset Creation}) and reported effect sizes (Tables~\ref{tab: maape hardness hypotheses}–\ref{tab: scc cross hypotheses}). The \emph{interaction of treatment and selection} is another threat, as knowledge efficacy may depend on system characteristics. Systems with high Module\# may benefit more from Complete knowledge. We included varied configurations and used statistical hypothesis testing (Section~\ref{sec: Methodology}), but synthetic data may not fully reflect real-world interactions.

\subsubsection{Conclusion}
A primary concern for our study's validity is external validity, which refers to the generalizability of our findings to real-world modular software systems. While our experiments used simulated models to achieve high control over independent variables, this approach inherently limits the realism of the settings compared to actual distributed systems. This tradeoff was a deliberate design choice. We prioritized high internal validity to establish clear causal relationships between structural aspects and performance modeling challenges. By using controlled simulations, we could systematically vary parameters and measure their effects with precision, avoiding the confounding variables common in real-world systems. We believe the value of this high internal validity outweighs the limitations in external validity, as it provides a robust foundation for understanding the underlying mechanisms. These insights can then guide future research and practical applications in more complex, real-world settings.

\section{Results And Discussions}
\label{sec: Results And Discussions}
This section provides the result of setups in Section~\ref{subsec: Addressing RQ1 Statistical Methods} and \ref{subsec: Addressing RQ2 Statistical Methods} to address RQ1 and RQ2, followed by a comprehensive discussion of the findings.

\subsection{Addressing RQ1: Influence of Modular System Aspects on Modeling Hardness}
\label{subsec: Addressing RQ1 in Results And Discussions}

\begin{figure}[htbp]
	
	\centering 
	
	\begin{subfigure}{0.49\textwidth}
		\includegraphics[width=\linewidth]{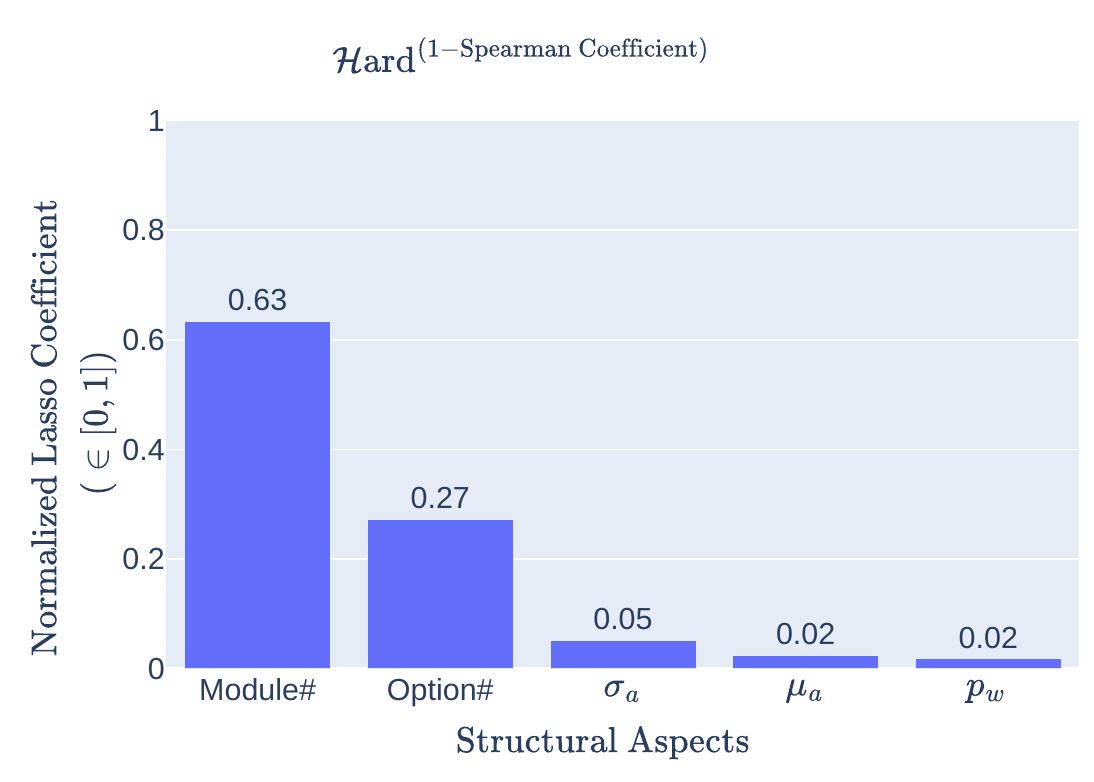}
		\caption{Hardness in terms of correlation}
		\label{fig:hardness analysis - corr}
	\end{subfigure}
	\hfill
	\begin{subfigure}{0.49\textwidth}
		\includegraphics[width=\linewidth]{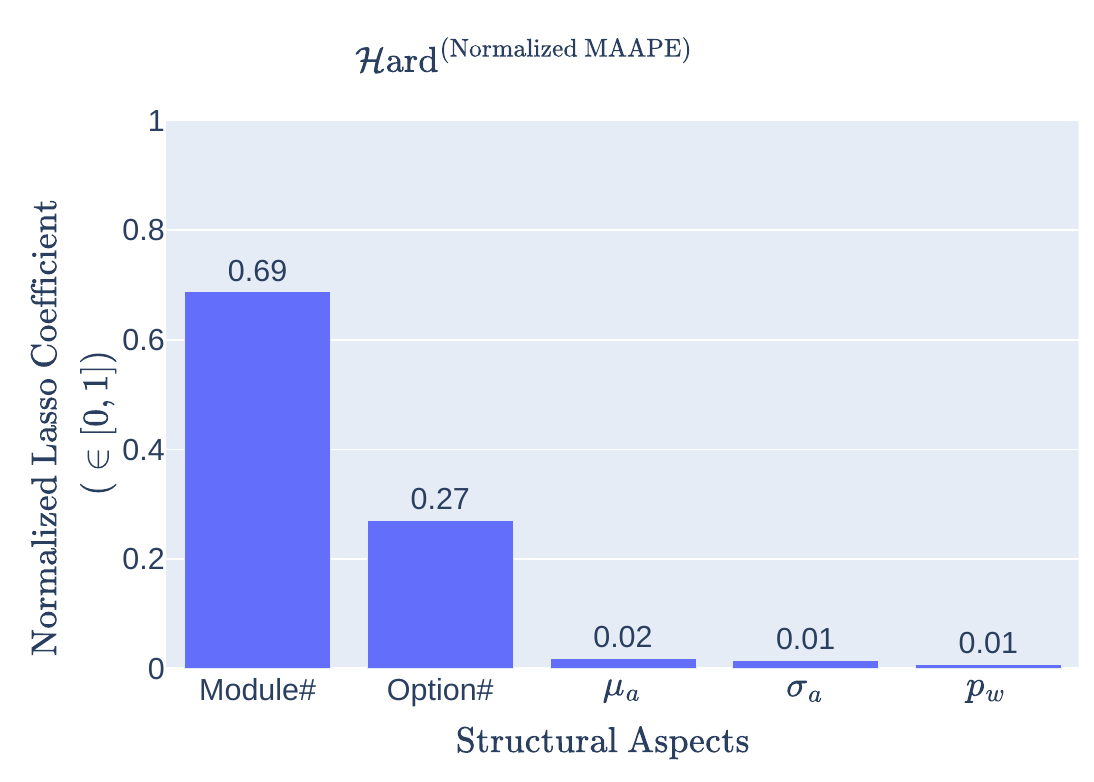}
		\caption{Hardness in terms of MAAPE}
		\label{fig:hardness analysis - MAAPE}
	\end{subfigure}
	\caption{Normalized weights of structural aspects (e.g., number of modules and configuration options, denoted Module\# and Option\# respectively) influencing performance modeling hardness in modular software systems, calculated using Lasso regression on 400 samples from synthetic modular systems. The graphs show how these aspects, particularly module count, drive modeling hardness, measured by ``$1 -\text{Spearman Coefficient}$'' (for ranking accuracy) and Normalized MAAPE loss (for prediction accuracy).}
	\label{fig: hardness analysis}
	
\end{figure}

Following the application of optimized regression techniques to the scaled dataset, as outlined in Section~\ref{subsec: Addressing RQ1 Statistical Methods}, the coefficients of the regression model variables were normalized to enhance interpretability. The results are depicted in Figure~\ref{fig: hardness analysis}.
\vspace{-15pt}

\vspace{-10pt}
\begin{figure}[!htbp]
	\centering
	\begin{subfigure}{0.49\textwidth}
		\centering
		\includegraphics[width=\linewidth]{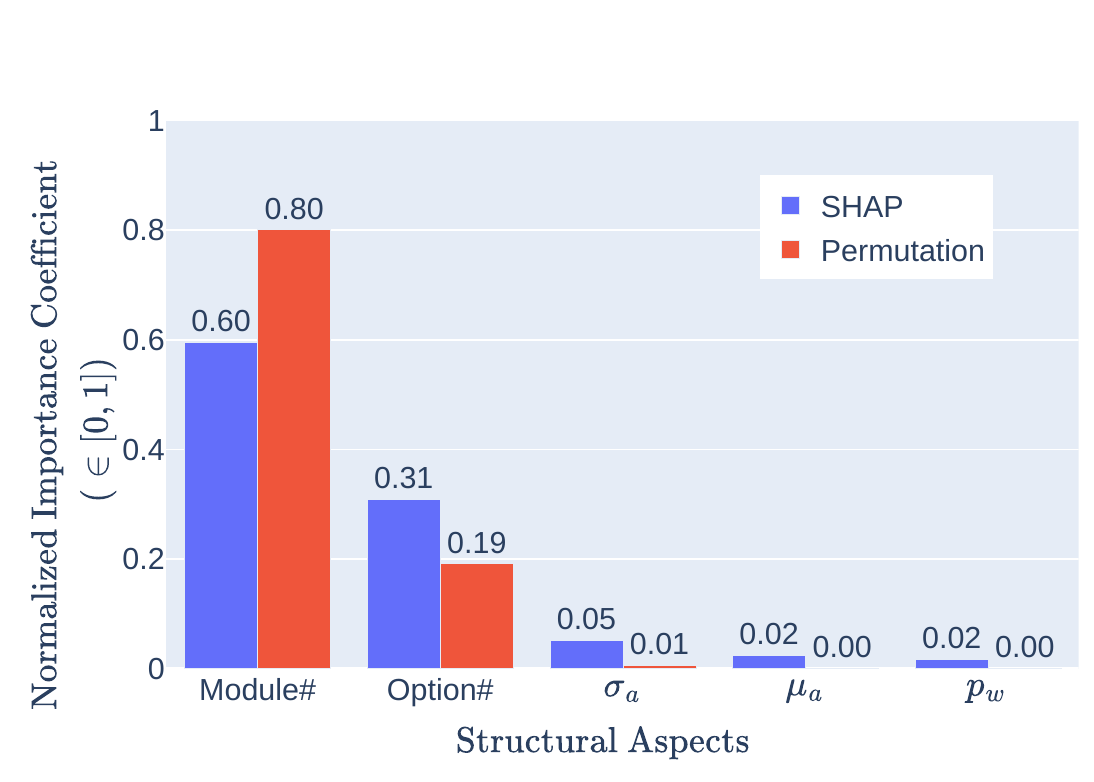}
		\caption{Normalized importance coefficients in hardness under ``$1 -\text{Spearman Coefficient}$'' loss metric, by SHAP and Permutation Importance.}
		\label{fig: importance hardness analysis - corr}
	\end{subfigure}
	\hfill
	\begin{subfigure}{0.49\textwidth}
		\centering
		\includegraphics[width=\linewidth]{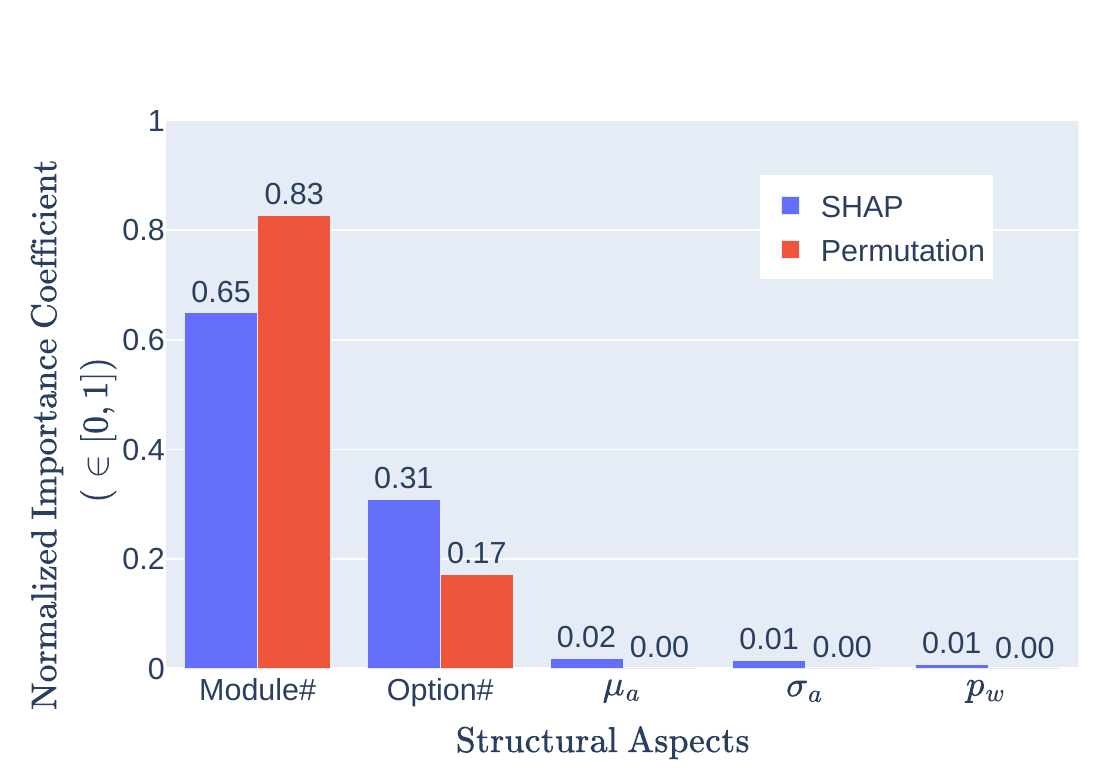}
		\caption{Normalized importance coefficients in hardness under Scaled MAAPE loss metric, by SHAP and Permutation Importance.}
		\label{fig: importance hardness analysis - MAAPE}
	\end{subfigure}
	
	\caption{Normalized weights of structural aspects (e.g., number of modules and configuration options) influencing performance modeling hardness in modular software systems, calculated using SHAP and Permutation Importance methods on 400 samples from synthetic modular systems. The graphs show how these aspects, particularly module count, drive modeling challenges, measured by ``$1 -\text{Spearman Coefficient}$'' (for ranking accuracy) and Normalized MAAPE loss (for prediction accuracy), highlighting the greater impact of module count on the effort needed for effective performance predictions using the black-box approach.}
	\label{fig: extra hardness analysis}
\end{figure}

Our analysis reveals that the system characteristics that make performance modeling most difficult are \emph{the number of modules} (Module\#) and \emph{the number of configuration options per module} (Option\#). 
However, the number of modules has a stronger impact on modeling hardness (above 0.6 out of 1), as observed from the highest coefficients in the regression model. 
This finding can be explained by the difference between \emph{local complexity} and \emph{structural complexity}. 
The number of options primarily contributes to local complexity within a single module, while the number of modules introduces a far more challenging structural complexity. 
As the number of modules increases, the potential for intricate inter-module interactions, dependencies, and communication overheads can grows exponentially. 
A performance model must not only learn the behavior of individual modules but also accurately capture this complex web of relationships.

For instance, consider a system with a total of 100 configuration options. A system with these 100 options distributed across just two modules (i.e., Option\# is 50 here) primarily requires the model to learn the performance characteristics of those two components and their single interaction. In contrast, if the same 100 options are spread across 20 modules (i.e., Option\# is 5 here), the model must now understand the performance of 20 individual components and, more importantly, the intricate network of relationships, dependencies, and communication patterns among them. This architectural complexity often poses a greater challenge for a model to learn and accurately predict than the combinatorial complexity of options within a few modules. 
Therefore, our results suggest that the hardness of modeling the interconnections and dependencies between modules is the dominant factor in determining overall modeling hardness.

While the probability of connections between configuration options and intermediate variables also play a role, their effect on modeling hardness is minimal compared to the number of modules and configuration options (in total, less than 0.1 out of 1).

To further validate these results, we conducted additional analyses using two importance metrics: SHAP (Shapley Additive Explanations)~\cite{lundberg2017unified} and Permutation Importance~\cite{breiman2001random}, represented in Figure~\ref{fig: extra hardness analysis}.
The findings from these tests were consistent with our previous results, with the impact of the number of modules scoring above 0.6 and 0.8 (out of 1) for SHAP and Permutation Importance, respectively, across both loss metrics. 
This consistency reinforces the reliability of our conclusions.

\begin{tcolorbox}[colback=gray!20, 
	colframe=gray!50, 
	boxrule=0.5mm, 
	arc=4mm, 
	width=\linewidth, 
	title={RQ1: How do changes in aspects of modular systems (Section~\ref{sec: structural aspects}) affect the hardness of performance modeling (Section~\ref{sec: hardness})?},
	fonttitle=\bfseries, 
	sharp corners=northwest] 
	
	Changes in modular system aspects, especially increasing the number of modules (Module\#) and configuration options per module (Option\#), make performance modeling harder, particularly with black-box approaches. 
	The number of modules has over twice the impact of configuration options due to broader structural complexity and interactions between components. 
	The ratio of causal connections between components have a smaller, indirect effect. 
	In practical terms, this means that for complex systems like microservices-based applications, black-box performance modeling is particularly hard, demanding significantly more effort and data to achieve effective performance predictions. 
	This makes tasks relying on performance-influence models, such as debugging, more time-consuming and resource-intensive, often necessitating advanced techniques or leveraging higher levels of structural knowledge for reliable results.
\end{tcolorbox}

\subsection{Addressing RQ2:  Impact of Structural Knowledge and Hardness on Modeling Opportunities}
\label{subsec: Addressing RQ2 in Results And Discussions}

This section addresses RQ2: How do different levels of structural knowledge (Partial, Practical, Complete) influence the opportunity to enhance performance modeling in modular software systems across varying levels of modeling hardness (Low, Medium, High)?

\begin{figure}[!htbp]
	\centering
	\begin{subfigure}{0.49\textwidth} 
		\centering
		\includegraphics[width=\linewidth]{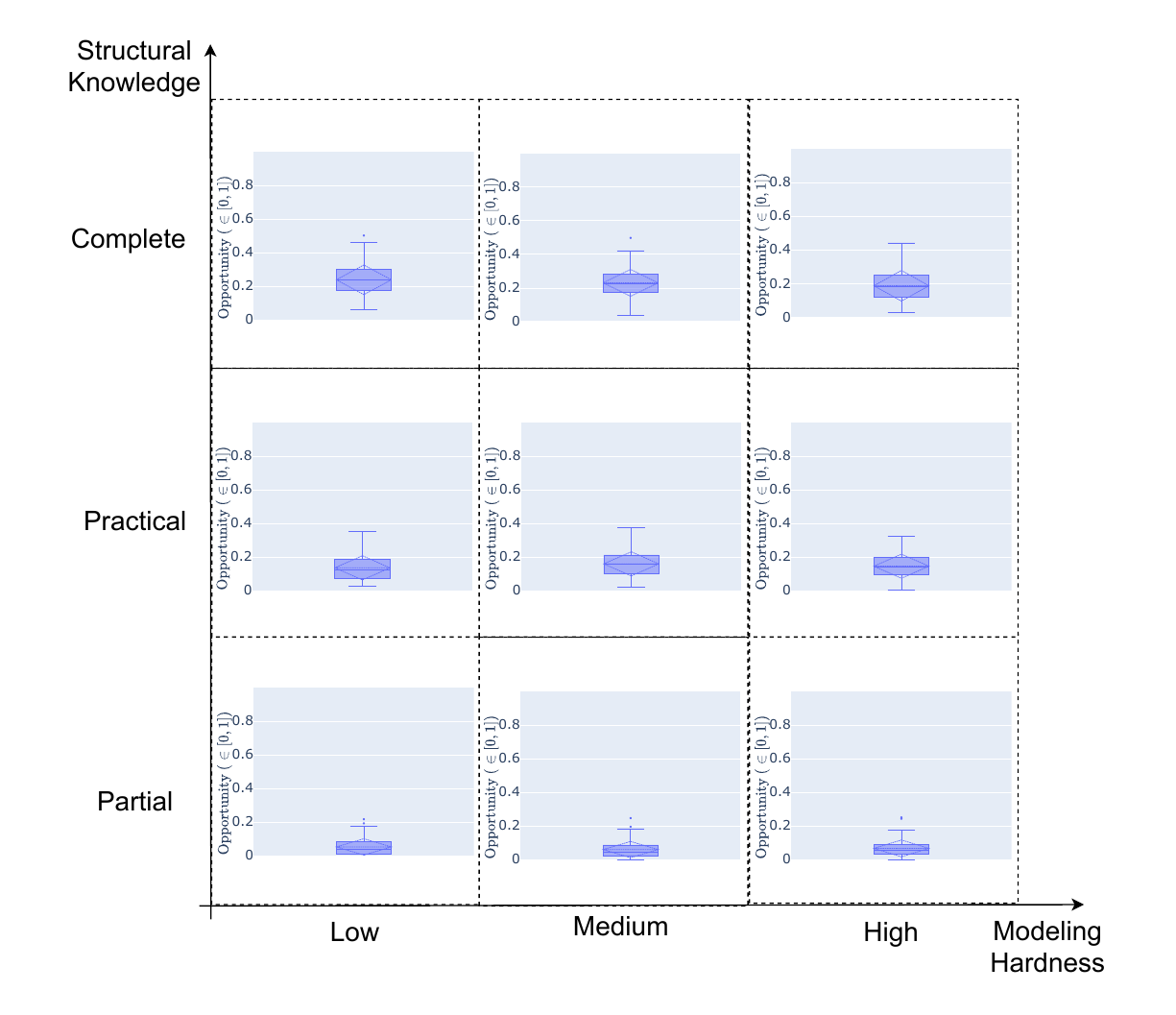}
		\caption{Analytical matrix for opportunities based on the correlation metric, i.e., $\avgopp^{(\mathrm{Spearman Coefficient},*,*)}$.}
		\label{fig: opportunity analysis - corr}
	\end{subfigure}
	\hfill
	\begin{subfigure}{0.49\textwidth} 
		\centering
		\includegraphics[width=\linewidth]{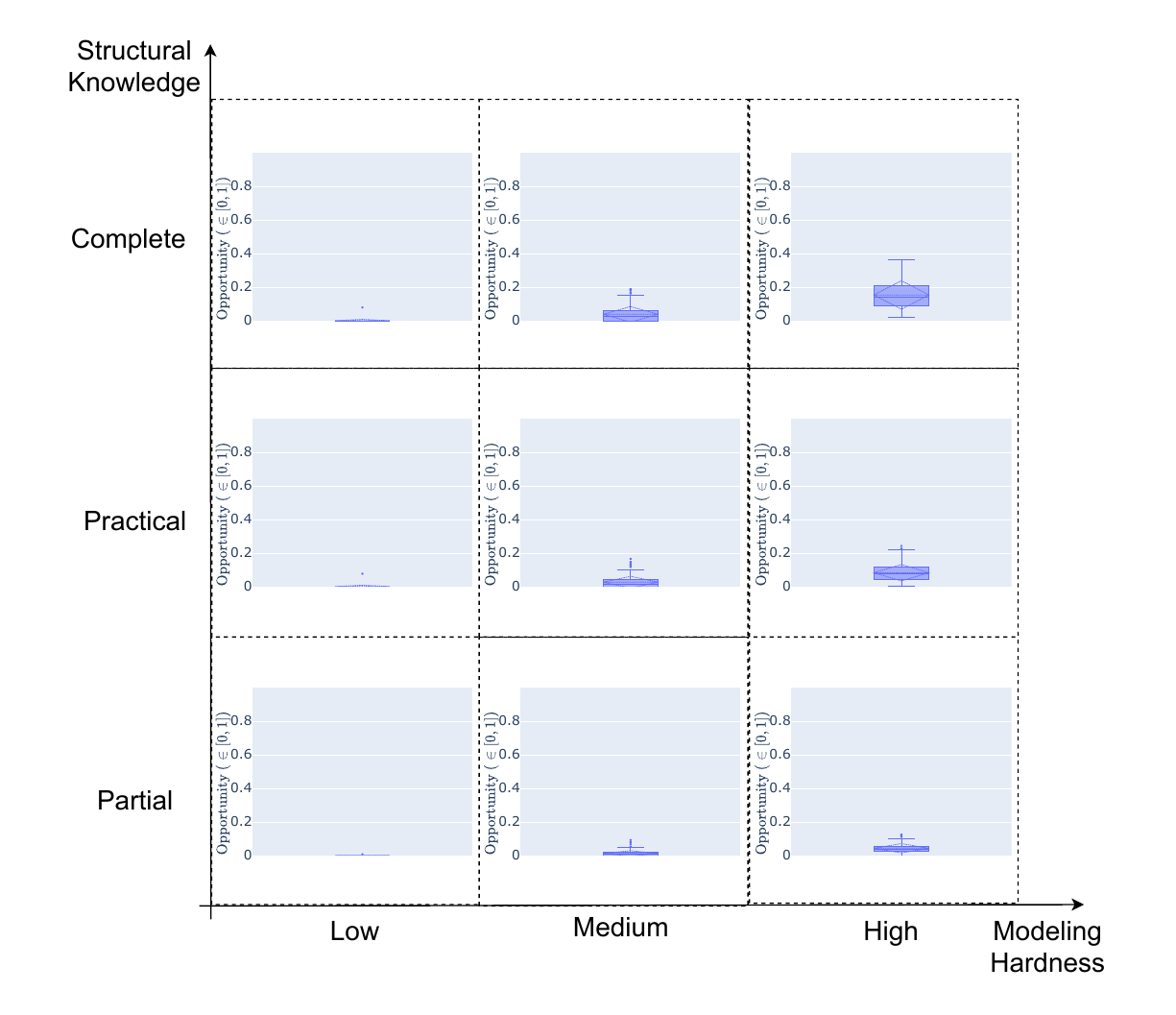}
		\caption{Analytical matrix for opportunities based on the accuracy efficacy metric, i.e., $\avgopp^{(\acc,*,*)}$.}
		\label{fig: opportunity analysis - MAAPE}
	\end{subfigure}
	
	\caption{Opportunity distribution based on Spearman Correlation Coefficient and $\acc$ efficacy metrics, presented in the form of the analytical matrix, represented in Figure~\ref{fig: raw analytical matrix}.}
	\label{fig: opportunity analysis distribution}
\end{figure}

To investigate this, we analyzed a dataset of 400 sampled modular systems, as detailed in Section~\ref{subsec: Synthtetic Dataset Creation}. Using the performance modeling approaches described in Section~\ref{subsec: Learning Methods for Performance Modeling}, we evaluated opportunity distributions for performance modeling. The results are visualized in Figure~\ref{fig: opportunity analysis distribution}, which presents box plots of opportunity levels for each system configuration, segmented by efficacy metrics (SCC for ranking accuracy and $\acc$ as a MAAPE-based metric for prediction accuracy).

To analyze the effect of modeling hardness on opportunity, we first categorized the calculated hardness values into three distinct levels (as described in Section~\ref{subsec: Addressing RQ2: Structural Knowledge and Hardness vs. Opportunity}): Low, Medium, and High.
This was achieved by discretizing the hardness values by splitting the data into four quartiles.
The first quartile was classified as Low hardness, the two middle quartiles were grouped together as Medium hardness, and the final quartile was designated as High hardness.
This classification method, inspired by the Interquartile Range (IQR) technique, provides a meaningful way to group our experimental results for comparative analysis.

To formally address RQ2, we conducted statistical hypothesis testing as outlined in Section~\ref{subsec: Addressing RQ2: Structural Knowledge and Hardness vs. Opportunity}, using methods described in Section~\ref{subsec: Addressing RQ1 Statistical Methods}. 
We performed 27 tests across three hypothesis groups: (1) hardness effects within fixed knowledge levels (horizontal comparisons in the analytical matrix, Figure~\ref{fig: raw analytical matrix}), (2) knowledge effects within fixed hardness levels (vertical comparisons), and (3) cross-comparisons between lower hardness with higher knowledge and higher hardness with lower knowledge.
Detailed results of these tests are provided in Appendix~\ref{appendix: Details of evaluation results}.

Our findings reveal that higher levels of structural knowledge consistently enhance the opportunity for improved performance modeling, with the magnitude of this effect varying based on the system's modeling hardness and the chosen efficacy metric. Specifically, systems with greater structural knowledge (e.g., Practical or Complete) demonstrate improved modeling efficacy, particularly for ranking accuracy (SCC).

\begin{figure}[!htbp]
	\centering
	\begin{subfigure}{0.49\textwidth} 
		\centering
		\includegraphics[width=\linewidth]{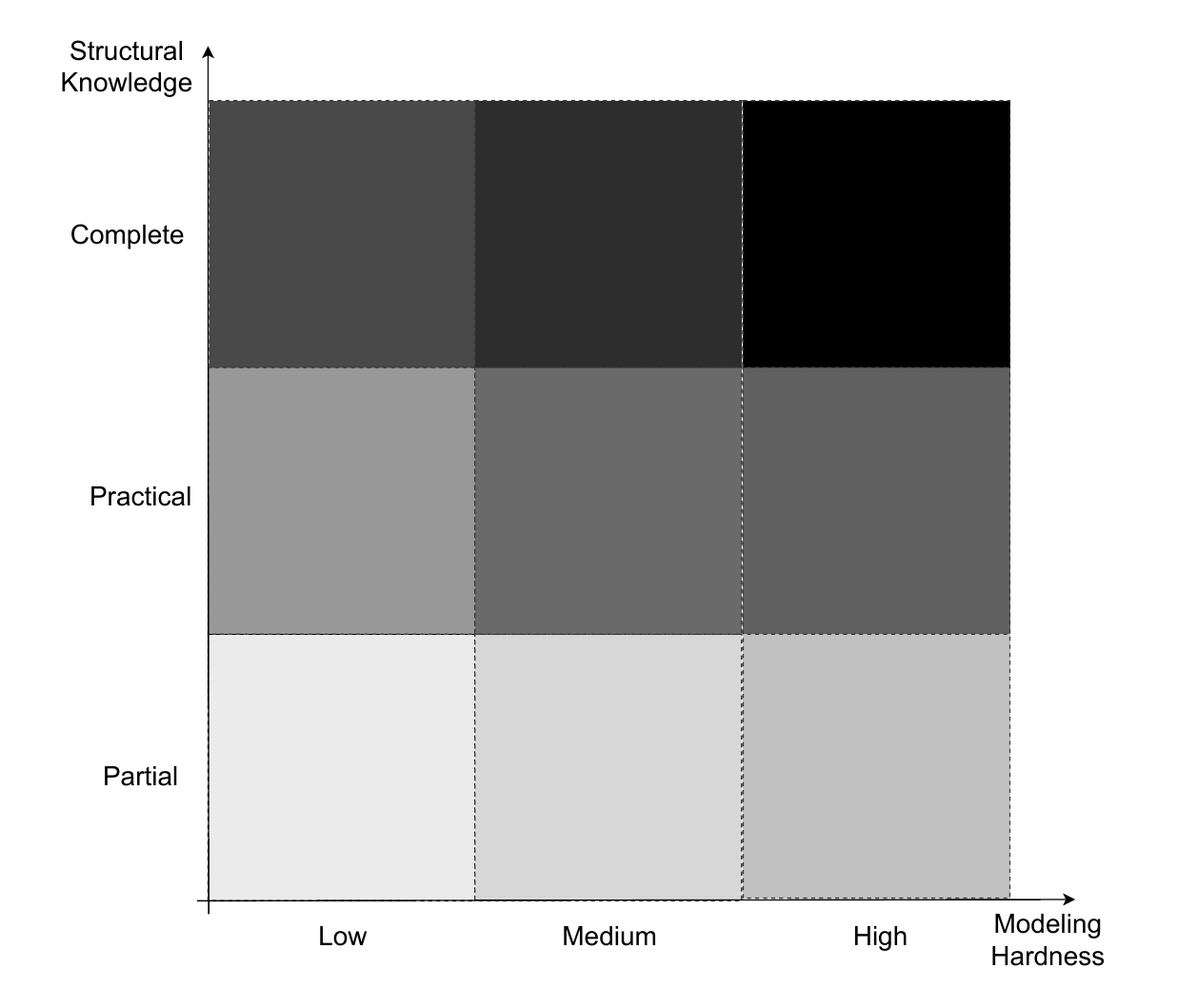}
		\caption{A heatmap visualization of the analytical matrix for opportunities based on the correlation metric, i.e., $\avgopp^{(\mathrm{Spearman Coefficient},*,*)}$.}
		\label{fig: heatmap opportunity analysis - corr}
	\end{subfigure}
	\hfill
	\begin{subfigure}{0.49\textwidth} 
		\centering
		\includegraphics[width=\linewidth]{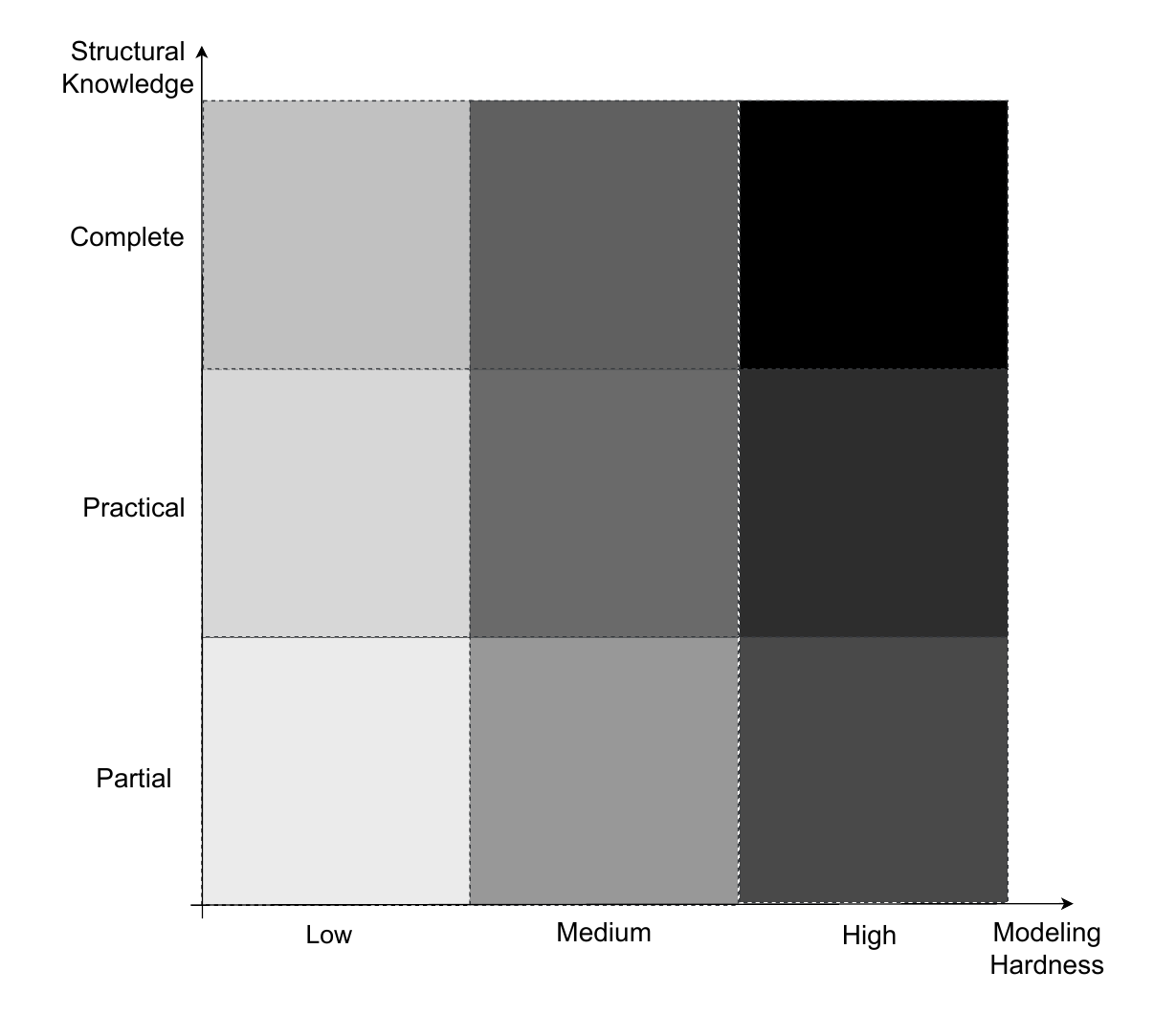}
		\caption{A heatmap visualization of the analytical matrix for opportunities based on the accuracy metric, i.e., $\avgopp^{(\acc,*,*)}$.}
		\label{fig: heatmap opportunity analysis - MAAPE}
	\end{subfigure}
	
	\caption{A heatmap visualization (bright to dark gradient) of opportunity distribution based on correlation- and accuracy-based metrics in the form of an analytical matrix, derived by statistical tests, which are explained in Section~\ref{subsec: Addressing RQ2: Structural Knowledge and Hardness vs. Opportunity}, on data represented in Figure~\ref{fig: opportunity analysis distribution}. Darker cells indicate higher expected opportunity levels.}
	\label{fig: heatmap opportunity analysis}
\end{figure}

This finding is further illustrated in Figure~\ref{fig: heatmap opportunity analysis}, which presents heatmaps using a bright-to-dark gradient to visualize opportunity distributions. The heatmaps highlight a stronger vertical (knowledge-driven) gradient for ranking accuracy scenarios (SCC), indicating that increased structural knowledge significantly boosts configuration ranking. Conversely, prediction accuracy scenarios (MAAPE) show a stronger horizontal (hardness-driven) gradient, with the greatest opportunities observed in high-hardness systems with Complete knowledge and the least in low-hardness systems with Partial knowledge.

\subsubsection*{\textbf{Key Insight 1: Structural Knowledge Boosts Ranking Accuracy (In SCC-based Scenarios) Across All Hardness Levels.}}

For scenarios where the goal is to accurately rank configurations by performance (e.g., identifying the fastest configurations in a system like the Hotel Reservation application), structural knowledge significantly enhances modeling opportunity. Statistical tests (Table~\ref{tab: scc kwnoledge hypotheses}) show that moving from Partial to Practical or Complete knowledge increases the opportunity to achieve higher SCC, with effect sizes ranging from $0.844$ to $0.968$ ($p < 0.05$). For example, at low hardness (simpler systems with fewer modules), Complete knowledge offers a $96.8$\,\% chance of outperforming Partial knowledge, meaning that knowing the full causal structure of a system dramatically improves ranking accuracy. 
This effect is slightly less pronounced at high hardness (e.g., $88.1$\,\% chance), where complex systems (i.e., with a higher level of modeling hardness, likely with many modules) require more data to model effectively, but the benefit of structural knowledge remains significant.

In practical terms, this suggests that system designers can improve configuration ranking--crucial for debugging-facing metrics like latency--by incorporating detailed knowledge of module boundaries and causal relationships, even in complex systems. 
For instance, in the Hotel Reservation application, Practical knowledge (including the execution graph) improved SCC from 0.178 to 0.192 (as expected from Figure~\ref{fig: heatmap opportunity analysis}), enabling better identification of low-latency configurations.

\subsubsection*{\textbf{Key Insight 2: Structural Knowledge Benefits Under Prediction Accuracy (In MAAPE-based Scenarios) Are Limited to Complex Systems.}}

When the goal is to minimize prediction errors (MAAPE), the impact of structural knowledge is less consistent. 
Statistical tests (Table~\ref{tab: maape knowledge hypotheses}) show no significant difference between Partial, Practical, and Complete knowledge at low and medium hardness levels ($p = 0.500$, effect size (CLES delta) $= 0.500$). This indicates that for simpler systems (i.e., lower modeling hardness), basic knowledge of module boundaries (Partial) is often sufficient, and additional knowledge does not substantially improve prediction accuracy. 
However, in high-hardness scenarios (complex systems with many modules), Complete knowledge begins to show benefits, though the effect is weaker than for SCC scenarios.

This finding highlights a trade-off: for systems where precise performance predictions are critical (e.g., ensuring latency meets strict thresholds), investing in detailed structural knowledge (e.g., practical level) is most valuable when dealing with highly complex systems. 
For example, in the Hotel Reservation application with the medium modeling hardness (explained in Section~\ref{subsec: Addressing RQ1 in Results And Discussions}), Practical knowledge could reduce prediction errors, but simpler systems may not justify the cost of acquiring such knowledge.

\subsubsection*{\textbf{Practical Implications.}}
These results guide system designers in prioritizing efforts to gather structural knowledge. For applications where ranking configurations (SCC) is critical--such as debugging microservices for user experience--investing in Practical or Complete knowledge (e.g., execution graphs or causal models) offers substantial benefits across all system complexities. 
For applications requiring precise predictions (MAAPE), such as meeting strict performance guarantees, designers should focus on detailed knowledge only for complex systems with many modules. These insights can inform resource allocation in real-world systems like those in the DeathStarBench suite, where understanding module interactions can streamline performance optimization.

\begin{tcolorbox}[colback=gray!20, 
	colframe=gray!50, 
	boxrule=0.5mm, 
	arc=4mm, 
	width=\linewidth, 
	title={RQ2: How do the levels of structural knowledge (Section~\ref{sec: structural knowledge}) and the modeling hardness impact the creation of opportunity (Section~\ref{sec: opportunity}) for improving modular performance modeling?},
	fonttitle=\bfseries, 
	sharp corners=northwest] 
	
	Higher structural knowledge consistently enhances the opportunity to improve performance modeling, particularly for ranking accuracy (SCC). 
	For prediction accuracy (MAAPE), knowledge effects are significant only in complex systems with high-hardness modeling. These findings underscore the value of structural knowledge in overcoming modeling challenges, offering actionable guidance for debugging and optimizing modular software systems.
\end{tcolorbox}

\subsection{Connection With The Motivating Observation}
This section examines how the observed discrepancy between the ``Self-care Mobile App'' and the ``Hotel Reservation'' application, as introduced in Section~\ref{sec: problem of hardness and opportunity}, aligns with the key findings presented in Section~\ref{subsec: Addressing RQ1 in Results And Discussions} and Section~\ref{subsec: Addressing RQ2 in Results And Discussions}.

First, we establish the modeling hardness levels for both applications.

\paragraph{Hotel Reservation Application} 
The modeling hardness of the Hotel Reservation application is classified as Medium. This is derived from a calculated hardness value of 0.718 out of 1, as shown in Figure~\ref{fig: Spearman Coefficnet of All For Hotel Reservation}. This value is obtained using the Spearman Correlation Coefficient across training data sizes $\{20,50,100,200,500,1000\}$ with the following formula:
\vspace{-0.5em}
\begin{equation*}
	\text{\footnotesize$
		\mathcal{H}ard^{(L)}(\mathcal{M}) = C^{(\mathcal{H}ard)} \times \left(\frac{(1-0.19)}{20} + \frac{(1-0.31)}{50} + \frac{(1-0.43)}{100} + \frac{(1-0.55)}{200} + \frac{(1-0.66)}{500} + \frac{(1-0.77)}{1000}\right)
		$,}
\end{equation*}

\noindent
where \(C^{(\mathcal{H}ard)} = \frac{1}{\sum_{i} \frac{1}{n_i}}\).
The application's structural complexity, characterized by 22 modules (Module\#) and at least 18 configuration options per module (Option\#), significantly contributes to this hardness. 
As supported by our findings (in Section~\ref{subsec: Addressing RQ1 in Results And Discussions}), the number of modules exerts a stronger influence on modeling hardness than the number of options, with regression weights of 0.63 and 0.27, respectively, for the ``$L := 1-\text{SCC}$'' metric. When compared to our synthetic systems, which range from 5 to 40 modules and 6 to 16 options (Table~\ref{tab: params for dataset}), the Hotel Reservation application's modeling hardness of 0.718 falls within the third quartile, confirming its classification as medium hardness.

\paragraph{Self-care Mobile App} 
In contrast, the Self-care Mobile App exhibits a Low modeling hardness. 
Its calculated hardness value is 0.201 out of 1 (Figure~\ref{fig: Spearman Coefficnet of All For Hotel Reservation}), derived from the Spearman Correlation Coefficient (SCC) using the following formula:
\vspace{-0.5em}
\begin{equation*}
	\text{\footnotesize$
		\mathcal{H}ard^{(L)}(\mathcal{M}) = C^{(\mathcal{H}ard)} \times \left(\frac{(1-0.71)}{20} + \frac{(1-0.87)}{50} + \frac{(1-0.96)}{100} + \frac{(1-0.97)}{200} + \frac{(1-0.97)}{500} + \frac{(1-0.98)}{1000}\right)
		$}.
\end{equation*}

\noindent
This application has a simpler structure, with only seven modules and six configuration options per module. Relative to the synthetic systems, its hardness value of 0.201 places it in the first quartile, consistent with a low hardness level.

\begin{figure}
	\centering
	\includegraphics[scale=0.7]{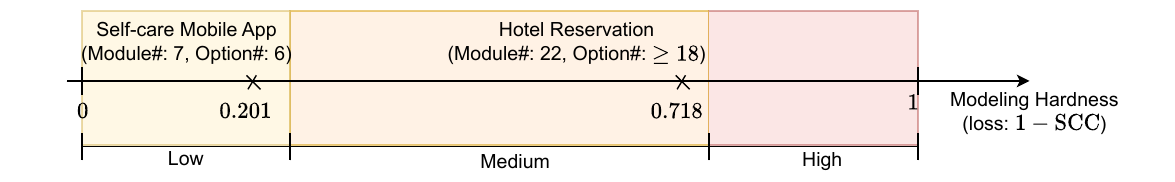}
	\caption{This figure represents the positioning of the applications discussed in Section~\ref{sec: problem of hardness and opportunity}, i.e., Self-care Mobile App and Hotel Reservation, mapping them onto the full spectrum of modeling hardness with respect to the loss metric $1-\text{SCC}$ (SCC is short for Spearman Correlation Coefficient), which ranges from 0 to 1. The figure provides a visual representation of how these applications fit within the hardness quartiles, which are defined as follows: the first quarter is classified as Low hardness, the two middle quarters as Medium, and the final quarter as High hardness.}
	\label{fig: motiv pos}
\end{figure}

Figure~\ref{fig: motiv pos} summarizes the positioning of these applications, illustrating how they fit within the modeling hardness spectrum from 0 to 1.

\paragraph{Alignment with Opportunity Analysis}
Based on these hardness levels and our key insights, we expect to see a greater opportunity for performance model improvement in the Hotel Reservation application compared to the Self-care App. 
This expectation is numerically supported by the opportunity values.
For the Hotel Reservation application, the opportunity at the Partial and Practical levels are 0.178 and 0.193, respectively (derived from Equation~\ref{eq: Opp}). 
For the Self-care App, these values are significantly lower, at 0.0 and 0.02. This result is in direct alignment with our first key insight and the final heatmap presented in Figure~\ref{fig: heatmap opportunity analysis - corr}.

\subsection{Future Work}
Future research should focus on developing \textit{module-wise sampling} techniques that exploit structural knowledge, such as distance-based metrics applied within individual modules. 
Unlike black-box distance sampling, which considers system-wide distances, \textit{module-wise sampling} accounts for intra-module distances, potentially improving modeling efficacy. 
Our preliminary observations indicate that adapting existing sampling techniques to this approach could enhance modeling opportunities, though this remains beyond the scope of the current study.

Another promising direction involves exploring the application of modular performance modeling under dynamic workloads in the context of self-adaptive systems. 
This includes investigating how performance models can evolve at runtime in response to dynamic structural knowledge, such as the addition or removal of modules. 
Maintaining up-to-date learning models is a critical challenge in machine-learning-enabled self-adaptive systems~\cite{gheibi2021applying} and represents a valuable area for future investigation.

Furthermore, integrating these research directions into the proposed analytical matrix will facilitate the identification of suitable methods for specific applications. Additional future work could explore alternative metrics, such as distribution accuracy metrics~\cite{chung2021beyond} (e.g., pinball loss), which are particularly relevant for risk-aware resource allocation, within the framework of the analytical matrix presented in this study.

\section{Conclusion}
\label{sec: conclusion}
This paper formally investigates the concepts of 
\emph{hardness} and \emph{opportunity} in modular performance modeling, bridging a critical gap in understanding how a system's structural characteristics and the level of available structural knowledge impact the efficacy of performance models. 
Through controlled experiments with synthetic system models (with a reproduction package at~\cite{ModularSystemSimulation}), the study establishes an analytical matrix to quantify these concepts. 
The key findings are that modeling hardness is driven primarily by the number of modules and configuration options per module, while causal connections have a negligible effect. 
Most importantly, the research demonstrates that both a higher level of structural knowledge and increased modeling hardness significantly enhance the opportunity for improvement in modular performance modeling. 
This indicates that acquiring and leveraging structural knowledge is most beneficial in complex, hard-to-model systems.

These results provide actionable insights for system designers, offering a guide on how to strategically allocate time and effort for modeling based on a system’s characteristics and the task's objectives. 
For instance, the paper reveals that structural knowledge is more dominant for improving ranking accuracy, a crucial metric for debugging tasks, whereas modeling hardness plays a stronger role for prediction accuracy, which is beneficial for tasks like resource management planning. 
Therefore, for systems that are inherently complex, designers should prioritize acquiring more detailed structural knowledge to effectively debug and improve performance. 
Conversely, simpler systems may achieve cost-effective performance modeling with less detailed knowledge, particularly for prediction-focused tasks. 
This work offers a systematic framework for understanding these trade-offs, enabling the selection of appropriate modeling approaches to achieve specific performance goals.

\begin{acks}
This work was partially supported by the \grantsponsor{NSF}{National Science Foundation}{https://www.nsf.gov/}
(Awards \grantnum{NSF}{2106853}, \grantnum{NSF}{2007202}, \grantnum{NSF}{2107463}, 
\grantnum{NSF}{2038080}, and \grantnum{NSF}{2233873}).
\end{acks}

\newpage
\appendix

\section{Details of Performance Modeling of Hotel Reservation Application}
\label{sec: Details of Performance Modeling of Hotel Reservation Application}
This section provides a detailed explanation of the configuration options, intermediate variables, discretization of the configuration space, and methods employed for performance modeling of the Hotel Reservation application, as referenced in Section~\ref{sec: problem of hardness and opportunity}.

\subsection{Configuration Options}
\label{app: Option Variables}

Table~\ref{tab: options for hotel reservation} outlines the configuration options for each module type in the Hotel Reservation application. The third column specifies the valid range of values for each option. However, due to the continuous nature of some options or their extensive value sets, evaluating the entire configuration space is impractical. To address this, we discretize these ranges into a manageable number of configurations, as indicated in the final column of the table.

\begin{table}[h!]
	\begin{center}
		\caption{Configuration options and their ranges for different module types in the Hotel Reservation application.}
		\label{tab: options for hotel reservation}
		\begin{tabular}{| >{\centering\arraybackslash}m{0.9in} | >{\centering\arraybackslash}m{1.6in} | >{\centering\arraybackslash}m{1.1in} | >{\centering\arraybackslash}m{1.1in} |}
			\hline
			\textbf{Module Type} & \textbf{Option Name}& \shortstack{\textbf{Valid Range of} \\ \textbf{Configuration} \\ \textbf{Space} } & \shortstack{\textbf{\# of} \\ \textbf{Configurations}}    \\ \hline
			
			\multirow{3}{*}{\shortstack{Go-based\\ Microservice} }  &  \multirow{1}{*}{\texttt{tcp\_rmem}} & \multirow{1}{*}{$\{4096..6291456\}$} & \multirow{1}{*}{6} \\  
			&    \multirow{1}{*}{\texttt{tcp\_wmem}} & \multirow{1}{*}{$\{4096..6291456\}$} & \multirow{1}{*}{6} \\
			&    \multirow{1}{*}{\texttt{cpu\_max\_ratio}} & \multirow{1}{*}{$[0,1]$} & \multirow{1}{*}{6} \\ \hline
			
			\multirow{3}{*}{\shortstack{Memcached} }  &  \multirow{1}{*}{\texttt{memory\_limit}} & \multirow{1}{*}{$\{4096..6291456\}$} & \multirow{1}{*}{6}\\  
			&    \multirow{1}{*}{\texttt{threads}} & \multirow{1}{*}{$\{1..16\}$} & \multirow{1}{*}{6}\\
			&    \multirow{1}{*}{\texttt{slab\_growth\_factor}} & \multirow{1}{*}{$[1.1,2.2]$} & \multirow{1}{*}{6}\\
			\hline
			
			\multirow{5}{*}{\shortstack{MongoDB} }  &  \multirow{1}{*}{\texttt{tcp\_rmem}} & \multirow{1}{*}{$\{4096..6291456\}$} & \multirow{1}{*}{6}\\  
			&    \multirow{1}{*}{\texttt{tcp\_wmem}} & \multirow{1}{*}{$\{4096..6291456\}$} & \multirow{1}{*}{6} \\
			&    \multirow{1}{*}{\texttt{eviction\_dirty\_target}} & \multirow{1}{*}{$\{10..90\}$} & \multirow{1}{*}{5} \\ 
			&    \multirow{1}{*}{\texttt{eviction\_dirty\_trigger}} & \multirow{1}{*}{$\{1..99\}$} & \multirow{1}{*}{6}\\
			&    \multirow{1}{*}{\texttt{wiredTigerCacheSizeGB}} & \multirow{1}{*}{$[0.05, 0.2]$} & \multirow{1}{*}{4}\\
			\hline
		\end{tabular}
	\end{center}
\end{table}

The total number of possible configurations is approximately $6^{66} \times 5^8 \times 4^8 \approx 2^{205}$. This estimate is derived as follows: for the 10 Go-based microservices, each with three options and six configurations per option, the total is $(6^3)^{10} = 6^{30}$; for the four Memcached services, each with three options and six configurations, the total is $(6^3)^4 = 6^{12}$; and for the eight MongoDB services, each with five options yielding $6^3 \times 5 \times 4$ configurations, the total is $(6^3 \times 5 \times 4)^8 = 6^{24} \times 5^8 \times 4^8$.

\subsection{Performance Modeling of the Hotel Reservation Application}
\label{subsec: Evaluation Setup For Performance Modeling Of Hotel Reservation}

This subsection elaborates on the setup and methods used to model the performance of the Hotel Reservation application.

To facilitate modeling, we discretize continuous configuration spaces. For instance, the range $[1.1, 2.2]$ for \texttt{slab\_growth\_factor} is divided into six evenly spaced values: $\{1.10$, $1.28$, $1.46$, $1.64, 1.82, 2.00\}$. Each discretized option is then one-hot encoded, transforming, for example, \texttt{slab\_growth\_factor} into six binary variables (e.g., \texttt{slab\_growth\_factor\_1.10}), where only one is active per configuration.

Random forest regression serves as our primary performance model, selected for its proven efficacy and interpretability in prior studies~\cite{grebhahn2019predicting, velez2020configcrusher, kaltenecker2020interplay}. 
In the Null and Ideal scenarios, we apply a standard random forest regressor. 
For the Partial scenario, we model each module individually with a random forest regressor, then aggregate their outputs using an additional random forest regressor to predict system-level performance. 
In the Practical scenario, we employ a structural causal model that incorporates potential causal relationships between options and intermediate variables. Irrelevant causal edges are pruned using the Fisher-Z test to have more performant model, and the remaining relationships—within and across modules—are modeled with random forest regressors, following approaches in~\cite{liang2023modular, iqbal2022unicorn}\footnote{Hyperparameters for the random forest regressors are tuned via Bayesian optimization to optimize model efficacy. Note that no single model is universally optimal across all scenarios\cite{grebhahn2019predicting}; efficacy depends on dataset characteristics and problem structure. Consistency in the learning method ensures comparability of evaluation results.}.

\subsection{Configuration Space Analysis}

\begin{figure}
	\centering
	\includegraphics[scale=0.4]{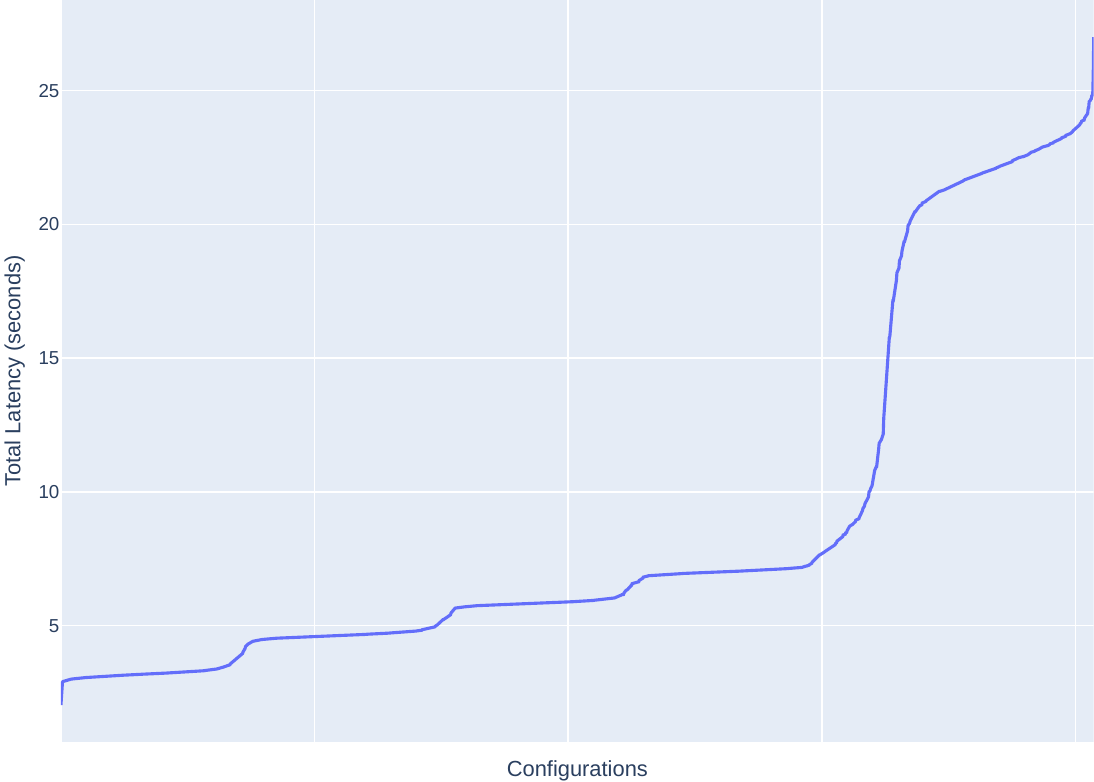}
	\caption{System latency for 2,000 configurations of the Hotel Reservation application, sorted from fastest to slowest.}
	\label{fig: hotel reservation performance behavior}
\end{figure}

To characterize system complexity, we measured latency across the 2,000 configurations (using \texttt{perf} tool and our own implemented package to automate configuring and running the application, which can be found in~\cite{DeathStarBenchIVsMeasurement}), and sorted them from fastest to slowest, as shown in Figure~\ref{fig: hotel reservation performance behavior}. While this analysis provides insights into system behavior~\cite{velez2021white}, it does not directly quantify the hardness of performance modeling.

\section{Details of evaluation results}
\label{appendix: Details of evaluation results}
This section presents the evaluation results discussed in Section~\ref{sec: Results And Discussions}, focusing on the analysis of modeling opportunities for modular software systems. 

\begin{figure}[!htbp]
	\centering
	\begin{subfigure}{0.49\textwidth} 
		\centering
		\includegraphics[width=\linewidth]{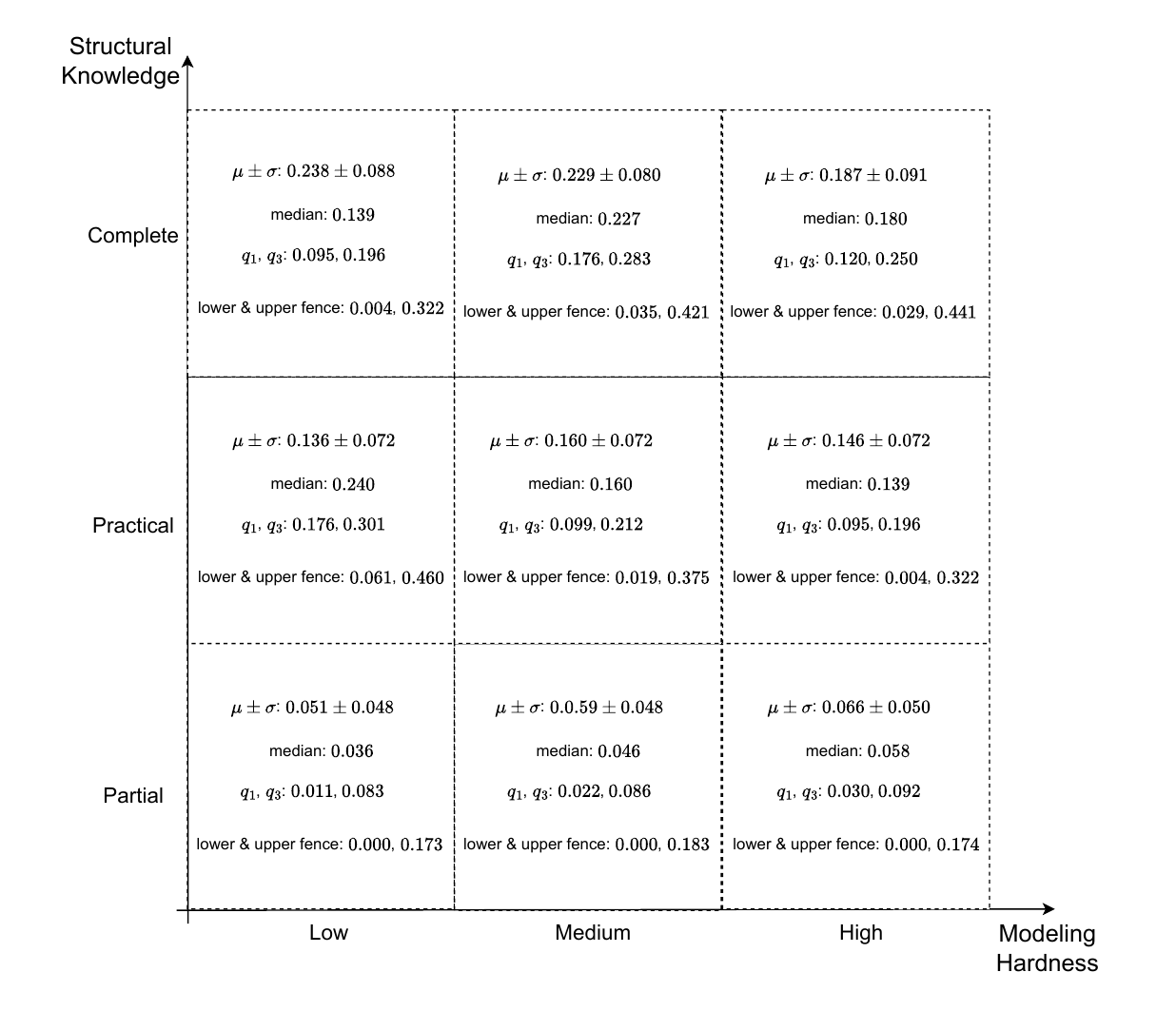}
		\caption{Analytical matrix for opportunities based on the correlation metric, i.e., $\avgopp^{(\mathrm{Spearman Coefficient},*,*)}$.}
		\label{fig: details of opportunity analysis - corr}
	\end{subfigure}
	\hfill
	\begin{subfigure}{0.49\textwidth} 
		\centering
		\includegraphics[width=\linewidth]{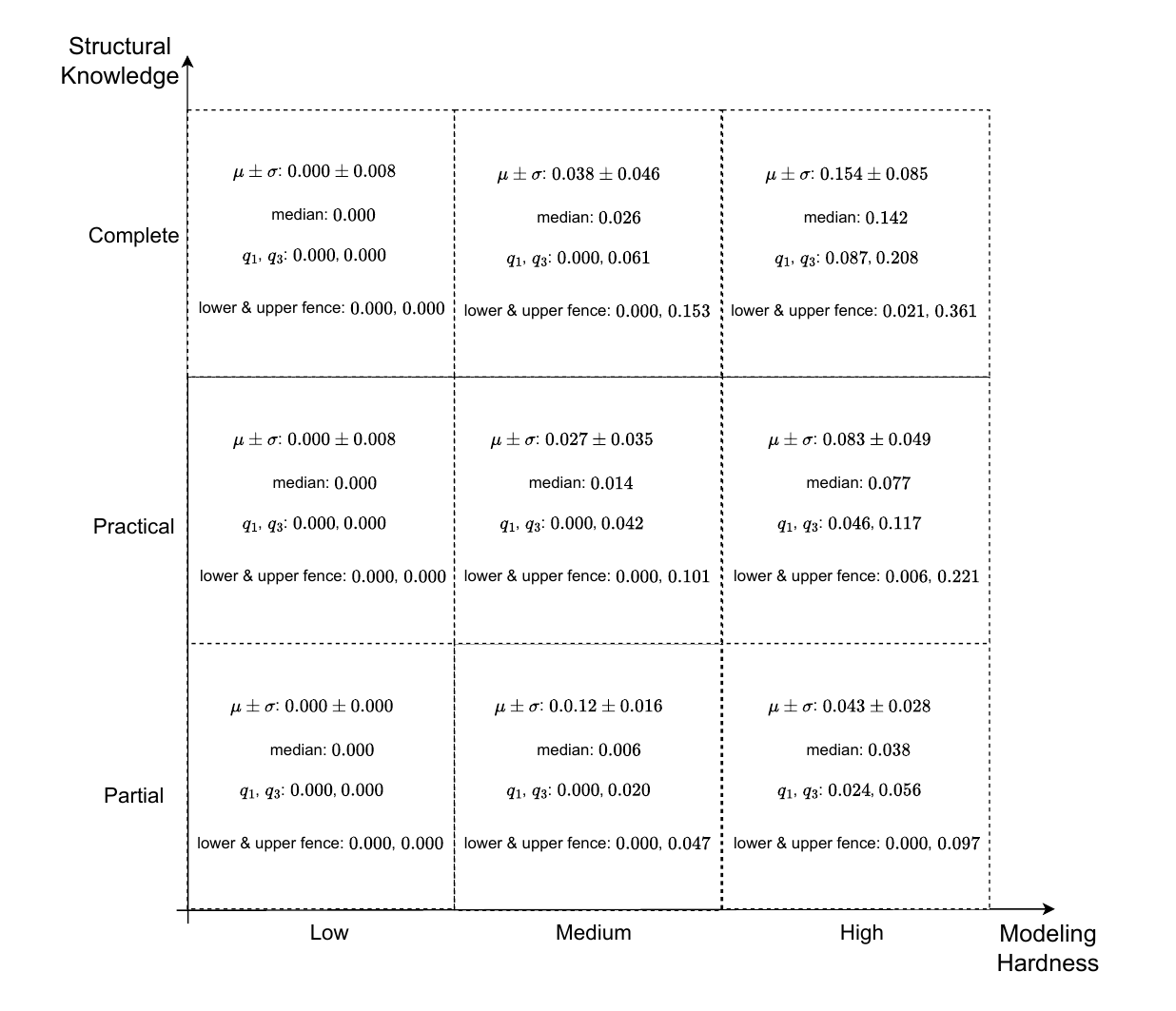}
		\caption{Analytical matrix for opportunities based on the accuracy efficacy metric, i.e., $\avgopp^{(\acc,*,*)}$.}
		\label{fig: details of opportunity analysis - MAAPE}
	\end{subfigure}

	\caption{Opportunity distribution statistics based on Spearman Coefficient Correlation and $\acc$ efficacy metrics in the form of the analytical matrix. These figures show the numerical information of distributions represented in Figure~\ref{fig: opportunity analysis distribution}.}
	\label{fig: opportunity analysis}
\end{figure}

Figure~\ref{fig: opportunity analysis} illustrates the detailed results of the opportunity analysis, summarizing the distribution statistics presented in Figure~\ref{fig: opportunity analysis distribution}.

The results of the statistical tests for all hypotheses related to the analytical matrix, evaluated using both efficacy metrics, are presented in detail in Tables~\ref{tab: maape hardness hypotheses} through~\ref{tab: scc cross hypotheses}. 

\begin{table}[h!]
	\begin{center}
		\caption{The statistical hypothesis testing results regarding the influence of the hardness level on the expected opportunity, as measured by the $\acc$ efficacy metric, under a fixed level of the structural knowledge.
			The last column reports the effect size for the group on the right-hand side of the null and alternative hypotheses.}
		\label{tab: maape hardness hypotheses}
		\begin{tabular}{|>{\centering\arraybackslash\small}m{3in}|*2{>{\centering\arraybackslash}m{0.7in}|} @{}m{0pt}@{}}
			\hline
			\shortstack{\textbf{Hypothesis}} & \shortstack{\textbf{p-value}}& \shortstack{\textbf{Effect Size} \\ \textbf{(CLES)} }  \\ \hline
			\multicolumn{3}{|c|}{Knowledge Level: Partial} \\ \hline
			\multirow{1}{*}{$H_0$: $\avgopp^{(\acc,partial,low)} = \avgopp^{(\acc,partial,medium)}$} & \multirow{2}{*}{\color{red}$0.000$} & \multirow{2}{*}{$0.770$} \\[5pt]
			\multirow{1}{*}{$H_a$: $\avgopp^{(\acc,partial,low)} < \avgopp^{(\acc,partial,medium)}$} & & \\[5pt]
			\hline
			\multirow{1}{*}{$H_0$: $\avgopp^{(\acc,partial,low)} = \avgopp^{(\acc,partial,high)}$} & \multirow{2}{*}{\color{red}$0.000$} & \multirow{2}{*}{$0.990$} \\[5pt]
			\multirow{1}{*}{$H_a$: $\avgopp^{(\acc,partial,low)} < \avgopp^{(\acc,partial,high)}$} & & \\[5pt]
			\hline
			\multirow{1}{*}{$H_0$: $\avgopp^{(\acc,partial,medium)} = \avgopp^{(\acc,partial,high)}$} & \multirow{2}{*}{\color{red}$0.000$} & \multirow{2}{*}{$0.861$} \\[5pt]
			\multirow{1}{*}{$H_a$: $\avgopp^{(\acc,partial,medium)} < \avgopp^{(\acc,partial,high)}$} & & \\[5pt]
			\hline
			\multicolumn{3}{|c|}{Knowledge Level: Practical} \\ \hline
			\multirow{1}{*}{$H_0$: $\avgopp^{(\acc,practical,low)} = \avgopp^{(\acc,practical,medium)}$} & \multirow{2}{*}{\color{red}$0.000$} & \multirow{2}{*}{$0.768$} \\[5pt]
			\multirow{1}{*}{$H_a$: $\avgopp^{(\acc,practical,low)} < \avgopp^{(\acc,practical,medium)}$} & & \\[5pt]
			\hline
			\multirow{1}{*}{$H_0$: $\avgopp^{(\acc,practical,low)} = \avgopp^{(\acc,practical,high)}$} & \multirow{2}{*}{\color{red}$0.000$} & \multirow{2}{*}{$0.995$} \\[5pt]
			\multirow{1}{*}{$H_a$: $\avgopp^{(\acc,practical,low)} < \avgopp^{(\acc,practical,high)}$} & & \\[5pt]
			\hline
			\multirow{1}{*}{$H_0$: $\avgopp^{(\acc,practical,medium)} = \avgopp^{(\acc,practical,high)}$} & \multirow{2}{*}{\color{red}$0.000$} & \multirow{2}{*}{$0.851$} \\[5pt]
			\multirow{1}{*}{$H_a$: $\avgopp^{(\acc,practical,medium)} < \avgopp^{(\acc,practical,high)}$} & & \\[5pt]
			\hline
			\multicolumn{3}{|c|}{Hardness Level: High} \\ \hline
			\multirow{1}{*}{$H_0$: $\avgopp^{(\acc,complete,low)} = \avgopp^{(\acc,complete,medium)}$} & \multirow{2}{*}{\color{red}$0.000$} & \multirow{2}{*}{$0.771$} \\[5pt]
			\multirow{1}{*}{$H_a$: $\avgopp^{(\acc,complete,low)} < \avgopp^{(\acc,complete,medium)}$} & & \\[5pt]
			\hline
			\multirow{1}{*}{$H_0$: $\avgopp^{(\acc,complete,low)} = \avgopp^{(\acc,complete,high)}$} & \multirow{2}{*}{\color{red}$0.000$} & \multirow{2}{*}{$0.998$} \\[5pt]
			\multirow{1}{*}{$H_a$: $\avgopp^{(\acc,complete,low)} < \avgopp^{(\acc,complete,high)}$} & & \\[5pt]
			\hline
			\multirow{1}{*}{$H_0$: $\avgopp^{(\acc,complete,medium)} = \avgopp^{(\acc,complete,high)}$} & \multirow{2}{*}{\color{red}$0.000$} & \multirow{2}{*}{$0.892$} \\[5pt]
			\multirow{1}{*}{$H_a$: $\avgopp^{(\acc,complete,medium)} < \avgopp^{(\acc,complete,high)}$} & & \\[5pt]
			\hline
		\end{tabular}
	\end{center}
\end{table}

\begin{table}[h!]
	\begin{center}
		\caption{The statistical hypothesis testing results regarding the influence of structural knowledge level on the expected opportunity, as measured by the $\acc$ efficacy metric, under a fixed level of the hardness.
			The last column reports the effect size for the group on the right-hand side of the null and alternative hypotheses.}
		\label{tab: maape knowledge hypotheses}
		\begin{tabular}{|>{\centering\arraybackslash\small}m{3in}|*2{>{\centering\arraybackslash}m{0.7in}|} @{}m{0pt}@{}}
			\hline
			\shortstack{\textbf{Hypothesis}} & \shortstack{\textbf{p-value}}& \shortstack{\textbf{Effect Size} \\ \textbf{(CLES)} }  \\ \hline
			\multicolumn{3}{|c|}{Hardness Level: Low} \\ \hline
			\multirow{1}{*}{$H_0$: $\avgopp^{(\acc,partial,low)} = \avgopp^{(\acc,practical,low)}$} & \multirow{2}{*}{$0.500$} & \multirow{2}{*}{$0.500$} \\[5pt]
			\multirow{1}{*}{$H_a$: $\avgopp^{(\acc,partial,low)} < \avgopp^{(\acc,practical,low)}$} & & \\[5pt]
			\hline
			\multirow{1}{*}{$H_0$: $\avgopp^{(\acc,partial,low)} = \avgopp^{(\acc,complete,low)}$} & \multirow{2}{*}{$0.500$} & \multirow{2}{*}{$0.500$} \\[5pt]
			\multirow{1}{*}{$H_a$: $\avgopp^{(\acc,partial,low)} < \avgopp^{(\acc,complete,low)}$} & & \\[5pt]
			\hline
			\multirow{1}{*}{$H_0$: $\avgopp^{(\acc,practical,low)} = \avgopp^{(\acc,complete,low)}$} & \multirow{2}{*}{$0.500$} & \multirow{2}{*}{$0.500$} \\[5pt]
			\multirow{1}{*}{$H_a$: $\avgopp^{(\acc,practical,low)} < \avgopp^{(\acc,complete,low)}$} & & \\[5pt]
			\hline
			\multicolumn{3}{|c|}{Hardness Level: Medium} \\ \hline
			\multirow{1}{*}{$H_0$: $\avgopp^{(\acc,partial,medium)} = \avgopp^{(\acc,practical,medium)}$} & \multirow{2}{*}{\color{red}$0.001$} & \multirow{2}{*}{$0.585$} \\[5pt]
			\multirow{1}{*}{$H_a$: $\avgopp^{(\acc,partial,medium)} < \avgopp^{(\acc,practical,medium)}$} & & \\[5pt]
			\hline
			\multirow{1}{*}{$H_0$: $\avgopp^{(\acc,partial,medium)} = \avgopp^{(\acc,complete,medium)}$} & \multirow{2}{*}{\color{red}$0.000$} & \multirow{2}{*}{$0.630$} \\[5pt]
			\multirow{1}{*}{$H_a$: $\avgopp^{(\acc,partial,medium)} < \avgopp^{(\acc,complete,medium)}$} & & \\[5pt]
			\hline
			\multirow{1}{*}{$H_0$: $\avgopp^{(\acc,practical,medium)} = \avgopp^{(\acc,complete,medium)}$} & \multirow{2}{*}{\color{red}$0.024$} & \multirow{2}{*}{$0.554$} \\[5pt]
			\multirow{1}{*}{$H_a$: $\avgopp^{(\acc,practical,medium)} < \avgopp^{(\acc,complete,medium)}$} & & \\[5pt]
			\hline
			\multicolumn{3}{|c|}{Hardness Level: High} \\ \hline
			\multirow{1}{*}{$H_0$: $\avgopp^{(\acc,partial,high)} = \avgopp^{(\acc,practical,high)}$} & \multirow{2}{*}{\color{red}$0.000$} & \multirow{2}{*}{$0.764$} \\[5pt]
			\multirow{1}{*}{$H_a$: $\avgopp^{(\acc,partial,high)} < \avgopp^{(\acc,practical,high)}$} & & \\[5pt]
			\hline
			\multirow{1}{*}{$H_0$: $\avgopp^{(\acc,partial,high)} = \avgopp^{(\acc,complete,high)}$} & \multirow{2}{*}{\color{red}$0.000$} & \multirow{2}{*}{$0.904$} \\[5pt]
			\multirow{1}{*}{$H_a$: $\avgopp^{(\acc,partial,high)} < \avgopp^{(\acc,complete,high)}$} & & \\[5pt]
			\hline
			\multirow{1}{*}{$H_0$: $\avgopp^{(\acc,practical,high)} = \avgopp^{(\acc,complete,high)}$} & \multirow{2}{*}{\color{red}$0.000$} & \multirow{2}{*}{$0.754$} \\[5pt]
			\multirow{1}{*}{$H_a$: $\avgopp^{(\acc,practical,high)} < \avgopp^{(\acc,complete,high)}$} & & \\[5pt]
			\hline
		\end{tabular}
	\end{center}
\end{table}

\begin{table}[h!]
	\begin{center}
		\caption{The statistical hypothesis testing results concerning the differential impact of hardness level versus structural knowledge level on the average opportunity—measured using the $\acc$ efficacy metric—are presented. The final two columns report the effect sizes for each data group, where G1 and G2 correspond to the groups on the left- and right-hand sides of the null hypothesis of equality, respectively.}
		\label{tab: maape cross hypotheses}
		\begin{adjustbox}{max width=\textwidth}
			\begin{tabular}{|>{\centering\arraybackslash\small}m{3in}|*3{>{\centering\arraybackslash}m{0.8in}|} @{}m{0pt}@{}}
				\hline
				\shortstack{\textbf{Hypothesis}} & \shortstack{\textbf{p-value}}& \shortstack{\footnotesize\textbf{Effect Size G1} \\ \textbf{(CLES)} } & \shortstack{\footnotesize\textbf{Effect Size G2} \\ \textbf{(CLES)} }  \\ \hline
				\multirow{1}{*}{$H_0$: $\avgopp^{(\acc,practical,low)} = \avgopp^{(\acc,partial,medium)}$} & \multirow{2}{*}{\color{red}$0.000$} & \multirow{2}{*}{$0.235$} & \multirow{2}{*}{\color{green}$0.765$} \\[5pt]
				\multirow{1}{*}{$H_a$: $\avgopp^{(\acc,practical,low)} \neq \avgopp^{(\acc,partial,medium)}$} & & &\\[5pt]
				\hline
				\multirow{1}{*}{$H_0$: $\avgopp^{(\acc,complete,low)} = \avgopp^{(\acc,partial,medium)}$} & \multirow{2}{*}{\color{red}$0.000$} & \multirow{2}{*}{$0.245$} & \multirow{2}{*}{\color{green}$0.765$} \\[5pt]
				\multirow{1}{*}{$H_a$: $\avgopp^{(\acc,complete,low)} \neq \avgopp^{(\acc,partial,medium)}$} & & &\\[5pt]
				\hline
				\multirow{1}{*}{$H_0$: $\avgopp^{(\acc,complete,low)} = \avgopp^{(\acc,practical,medium)}$} & \multirow{2}{*}{\color{red}$0.000$} & \multirow{2}{*}{$0.232$} & \multirow{2}{*}{\color{green}$0.768$} \\[5pt]
				\multirow{1}{*}{$H_a$: $\avgopp^{(\acc,complete,low)} \neq \avgopp^{(\acc,practical,medium)}$} & & & \\[5pt]
				\hline
				\multirow{1}{*}{$H_0$: $\avgopp^{(\acc,practical,low)} = \avgopp^{(\acc,partial,high)}$} & \multirow{2}{*}{\color{red}$0.000$} & \multirow{2}{*}{$0.091$} & \multirow{2}{*}{\color{green}$0.981$} \\[5pt]
				\multirow{1}{*}{$H_a$: $\avgopp^{(\acc,practical,low)} \neq \avgopp^{(\acc,partial,high)}$} & & & \\[5pt]
				\hline
				\multirow{1}{*}{$H_0$: $\avgopp^{(\acc,complete,low)} = \avgopp^{(\acc,partial,high)}$} & \multirow{2}{*}{\color{red}$0.000$} & \multirow{2}{*}{$0.019$} & \multirow{2}{*}{\color{green}$0.981$} \\[5pt]
				\multirow{1}{*}{$H_a$: $\avgopp^{(\acc,complete,low)} \neq \avgopp^{(\acc,partial,high)}$} & & & \\[5pt]
				\hline
				\multirow{1}{*}{$H_0$: $\avgopp^{(\acc,practical,medium)} = \avgopp^{(\acc,partial,high)}$} & \multirow{2}{*}{\color{red}$0.000$} & \multirow{2}{*}{$0.295$} & \multirow{2}{*}{\color{green}$0.705$} \\[5pt]
				\multirow{1}{*}{$H_a$: $\avgopp^{(\acc,practical,medium)} \neq \avgopp^{(\acc,partial,high)}$} & & & \\[5pt]
				\hline
				\multirow{1}{*}{$H_0$: $\avgopp^{(\acc,complete,medium)} = \avgopp^{(\acc,partial,high)}$} & \multirow{2}{*}{\color{red}$0.000$} & \multirow{2}{*}{$0.399$} & \multirow{2}{*}{\color{green}$0.601$} \\[5pt]
				\multirow{1}{*}{$H_a$: $\avgopp^{(\acc,complete,medium)} \neq \avgopp^{(\acc,partial,high)}$} & & &\\[5pt]
				\hline
				\multirow{1}{*}{$H_0$: $\avgopp^{(\acc,complete,low)} = \avgopp^{(\acc,practical,high)}$} & \multirow{2}{*}{\color{red}$0.000$} & \multirow{2}{*}{$0.005$} & \multirow{2}{*}{\color{green}$0.995$} \\[5pt]
				\multirow{1}{*}{$H_a$: $\avgopp^{(\acc,complete,low)} \neq \avgopp^{(\acc,practical,high)}$} & & &\\[5pt]
				\hline
				\multirow{1}{*}{$H_0$: $\avgopp^{(\acc,complete,medium)} = \avgopp^{(\acc,practical,high)}$} & \multirow{2}{*}{\color{red}$0.000$} & \multirow{2}{*}{$0.229$} & \multirow{2}{*}{\color{green}$0.771$} \\[5pt]
				\multirow{1}{*}{$H_a$: $\avgopp^{(\acc,complete,medium)} \neq \avgopp^{(\acc,practical,high)}$} & & & \\[5pt]
				\hline
			\end{tabular}
		\end{adjustbox}
	\end{center}
\end{table}

\begin{table}[h!]
	\begin{center}
		\caption{The statistical hypothesis testing results regarding the influence of the hardness level on the expected opportunity, as measured by the SCC efficacy metric, under a fixed level of the structural knowledge.
			The last column reports the effect size for the group on the right-hand side of the null and alternative hypotheses.}
		\label{tab: scc hardness hypotheses}
		\begin{tabular}{|>{\centering\arraybackslash\small}m{3in}|*2{>{\centering\arraybackslash}m{0.7in}|} @{}m{0pt}@{}}
			\hline
			\shortstack{\textbf{Hypothesis}} & \shortstack{\textbf{p-value}}& \shortstack{\textbf{Effect Size} \\ \textbf{(CLES)} }  \\ \hline
			\multicolumn{3}{|c|}{Knowledge Level: Partial} \\ \hline
			\multirow{1}{*}{$H_0$: $\avgopp^{(\mathrm{SCC},partial,low)} = \avgopp^{(\mathrm{SCC},partial,medium)}$} & \multirow{2}{*}{\color{red}$0.037$} & \multirow{2}{*}{$0.563$} \\[5pt]
			\multirow{1}{*}{$H_a$: $\avgopp^{(\mathrm{SCC},partial,low)} < \avgopp^{(\mathrm{SCC},partial,medium)}$} & & \\[5pt]
			\hline
			\multirow{1}{*}{$H_0$: $\avgopp^{(\mathrm{SCC},partial,low)} = \avgopp^{(\mathrm{SCC},partial,high)}$} & \multirow{2}{*}{\color{red}$0.005$} & \multirow{2}{*}{$0.606$} \\[5pt]
			\multirow{1}{*}{$H_a$: $\avgopp^{(\mathrm{SCC},partial,low)} < \avgopp^{(\mathrm{SCC},partial,high)}$} & & \\[5pt]
			\hline
			\multirow{1}{*}{$H_0$: $\avgopp^{(\mathrm{SCC},partial,medium)} = \avgopp^{(\mathrm{SCC},partial,high)}$} & \multirow{2}{*}{$0.100$} & \multirow{2}{*}{$0.545$} \\[5pt]
			\multirow{1}{*}{$H_a$: $\avgopp^{(\mathrm{SCC},partial,medium)} < \avgopp^{(\mathrm{SCC},partial,high)}$} & & \\[5pt]
			\hline
			\multicolumn{3}{|c|}{Knowledge Level: Practical} \\ \hline
			\multirow{1}{*}{$H_0$: $\avgopp^{(\mathrm{SCC},practical,low)} = \avgopp^{(\mathrm{SCC},practical,medium)}$} & \multirow{2}{*}{\color{red}$0.002$} & \multirow{2}{*}{$0.601$} \\[5pt]
			\multirow{1}{*}{$H_a$: $\avgopp^{(\mathrm{SCC},practical,low)} < \avgopp^{(\mathrm{SCC},practical,medium)}$} & & \\[5pt]
			\hline
			\multirow{1}{*}{$H_0$: $\avgopp^{(\mathrm{SCC},practical,low)} = \avgopp^{(\mathrm{SCC},practical,high)}$} & \multirow{2}{*}{$0.148$} & \multirow{2}{*}{$0.543$} \\[5pt]
			\multirow{1}{*}{$H_a$: $\avgopp^{(\mathrm{SCC},practical,low)} < \avgopp^{(\mathrm{SCC},practical,high)}$} & & \\[5pt]
			\hline
			\multirow{1}{*}{$H_0$: $\avgopp^{(\mathrm{SCC},practical,medium)} = \avgopp^{(\mathrm{SCC},practical,high)}$} & \multirow{2}{*}{$0.944$} & \multirow{2}{*}{$0.444$} \\[5pt]
			\multirow{1}{*}{$H_a$: $\avgopp^{(\mathrm{SCC},practical,medium)} < \avgopp^{(\mathrm{SCC},practical,high)}$} & & \\[5pt]
			\hline
			\multicolumn{3}{|c|}{Hardness Level: High} \\ \hline
			\multirow{1}{*}{$H_0$: $\avgopp^{(\mathrm{SCC},complete,low)} = \avgopp^{(\mathrm{SCC},complete,medium)}$} & \multirow{2}{*}{$0.767$} & \multirow{2}{*}{$0.474$} \\[5pt]
			\multirow{1}{*}{$H_a$: $\avgopp^{(\mathrm{SCC},complete,low)} < \avgopp^{(\mathrm{SCC},complete,medium)}$} & & \\[5pt]
			\hline
			\multirow{1}{*}{$H_0$: $\avgopp^{(\mathrm{SCC},complete,low)} = \avgopp^{(\mathrm{SCC},complete,high)}$} & \multirow{2}{*}{$1.000$} & \multirow{2}{*}{$0.341$} \\[5pt]
			\multirow{1}{*}{$H_a$: $\avgopp^{(\mathrm{SCC},complete,low)} < \avgopp^{(\mathrm{SCC},complete,high)}$} & & \\[5pt]
			\hline
			\multirow{1}{*}{$H_0$: $\avgopp^{(\mathrm{SCC},complete,medium)} = \avgopp^{(\mathrm{SCC},complete,high)}$} & \multirow{2}{*}{$1.000$} & \multirow{2}{*}{$0.359$} \\[5pt]
			\multirow{1}{*}{$H_a$: $\avgopp^{(\mathrm{SCC},complete,medium)} < \avgopp^{(\mathrm{SCC},complete,high)}$} & & \\[5pt]
			\hline
		\end{tabular}
	\end{center}
\end{table}

\begin{table}[h!]
	\begin{center}
		\caption{The statistical hypothesis testing results regarding the influence of structural knowledge level on the expected opportunity, as measured by the SCC efficacy metric, under a fixed level of the hardness.
			The last column reports the effect size for the group on the right-hand side of the null and alternative hypotheses.}
		\label{tab: scc kwnoledge hypotheses}
		\begin{tabular}{|>{\centering\arraybackslash\small}m{3in}|*2{>{\centering\arraybackslash}m{0.7in}|} @{}m{0pt}@{}}
			\hline
			\shortstack{\textbf{Hypothesis}} & \shortstack{\textbf{p-value}}& \shortstack{\textbf{Effect Size} \\ \textbf{(CLES)} }  \\ \hline
			\multicolumn{3}{|c|}{Hardness Level: Low} \\ \hline
			\multirow{1}{*}{$H_0$: $\avgopp^{(\mathrm{SCC},partial,low)} = \avgopp^{(\mathrm{SCC},practical,low)}$} & \multirow{2}{*}{\color{red}$0.000$} & \multirow{2}{*}{$0.844$} \\[5pt]
			\multirow{1}{*}{$H_a$: $\avgopp^{(\mathrm{SCC},partial,low)} < \avgopp^{(\mathrm{SCC},practical,low)}$} & & \\[5pt]
			\hline
			\multirow{1}{*}{$H_0$: $\avgopp^{(\mathrm{SCC},partial,low)} = \avgopp^{(\mathrm{SCC},complete,low)}$} & \multirow{2}{*}{\color{red}$0.000$} & \multirow{2}{*}{$0.968$} \\[5pt]
			\multirow{1}{*}{$H_a$: $\avgopp^{(\mathrm{SCC},partial,low)} < \avgopp^{(\mathrm{SCC},complete,low)}$} & & \\[5pt]
			\hline
			\multirow{1}{*}{$H_0$: $\avgopp^{(\mathrm{SCC},practical,low)} = \avgopp^{(\mathrm{SCC},complete,low)}$} & \multirow{2}{*}{\color{red}$0.000$} & \multirow{2}{*}{$0.812$} \\[5pt]
			\multirow{1}{*}{$H_a$: $\avgopp^{(\mathrm{SCC},practical,low)} < \avgopp^{(\mathrm{SCC},complete,low)}$} & & \\[5pt]
			\hline
			\multicolumn{3}{|c|}{Hardness Level: Medium} \\ \hline
			\multirow{1}{*}{$H_0$: $\avgopp^{(\mathrm{SCC},partial,medium)} = \avgopp^{(\mathrm{SCC},practical,medium)}$} & \multirow{2}{*}{\color{red}$0.000$} & \multirow{2}{*}{$0.877$} \\[5pt]
			\multirow{1}{*}{$H_a$: $\avgopp^{(\mathrm{SCC},partial,medium)} < \avgopp^{(\mathrm{SCC},practical,medium)}$} & & \\[5pt]
			\hline
			\multirow{1}{*}{$H_0$: $\avgopp^{(\mathrm{SCC},partial,medium)} = \avgopp^{(\mathrm{SCC},complete,medium)}$} & \multirow{2}{*}{\color{red}$0.000$} & \multirow{2}{*}{$0.964$} \\[5pt]
			\multirow{1}{*}{$H_a$: $\avgopp^{(\mathrm{SCC},partial,medium)} < \avgopp^{(\mathrm{SCC},complete,medium)}$} & & \\[5pt]
			\hline
			\multirow{1}{*}{$H_0$: $\avgopp^{(\mathrm{SCC},practical,medium)} = \avgopp^{(\mathrm{SCC},complete,medium)}$} & \multirow{2}{*}{\color{red}$0.000$} & \multirow{2}{*}{$0.738$} \\[5pt]
			\multirow{1}{*}{$H_a$: $\avgopp^{(\mathrm{SCC},practical,medium)} < \avgopp^{(\mathrm{SCC},complete,medium)}$} & & \\[5pt]
			\hline
			\multicolumn{3}{|c|}{Hardness Level: High} \\ \hline
			\multirow{1}{*}{$H_0$: $\avgopp^{(\mathrm{SCC},partial,high)} = \avgopp^{(\mathrm{SCC},practical,high)}$} & \multirow{2}{*}{\color{red}$0.000$} & \multirow{2}{*}{$0.822$} \\[5pt]
			\multirow{1}{*}{$H_a$: $\avgopp^{(\mathrm{SCC},partial,high)} < \avgopp^{(\mathrm{SCC},practical,high)}$} & & \\[5pt]
			\hline
			\multirow{1}{*}{$H_0$: $\avgopp^{(\mathrm{SCC},partial,high)} = \avgopp^{(\mathrm{SCC},complete,high)}$} & \multirow{2}{*}{\color{red}$0.000$} & \multirow{2}{*}{$0.881$} \\[5pt]
			\multirow{1}{*}{$H_a$: $\avgopp^{(\mathrm{SCC},partial,high)} < \avgopp^{(\mathrm{SCC},complete,high)}$} & & \\[5pt]
			\hline
			\multirow{1}{*}{$H_0$: $\avgopp^{(\mathrm{SCC},practical,high)} = \avgopp^{(\mathrm{SCC},complete,high)}$} & \multirow{2}{*}{\color{red}$0.001$} & \multirow{2}{*}{$0.623$} \\[5pt]
			\multirow{1}{*}{$H_a$: $\avgopp^{(\mathrm{SCC},practical,high)} < \avgopp^{(\mathrm{SCC},complete,high)}$} & & \\[5pt]
			\hline
		\end{tabular}
	\end{center}
\end{table}

\begin{table}[h!]
	\begin{center}
		\caption{The statistical hypothesis testing results concerning the differential impact of hardness level versus structural knowledge level on the average opportunity—measured using the SCC efficacy metric—are presented. The final two columns report the effect sizes for each data group, where G1 and G2 correspond to the groups on the left- and right-hand sides of the null hypothesis of equality, respectively.}
		\label{tab: scc cross hypotheses}
		\begin{adjustbox}{max width=\textwidth}
			\begin{tabular}{|>{\centering\arraybackslash\small}m{3in}|*3{>{\centering\arraybackslash}m{0.8in}|} @{}m{0pt}@{}}
				\hline
				\shortstack{\textbf{Hypothesis}} & \shortstack{\textbf{p-value}}& \shortstack{\footnotesize\textbf{Effect Size G1} \\ \textbf{(CLES)} } & \shortstack{\footnotesize\textbf{Effect Size G2} \\ \textbf{(CLES)} }  \\ \hline
				\multirow{1}{*}{$H_0$: $\avgopp^{(\mathrm{SCC},practical,low)} = \avgopp^{(\mathrm{SCC},partial,medium)}$} & \multirow{2}{*}{\color{red}$0.000$} & \multirow{2}{*}{\color{green}$0.816$} & \multirow{2}{*}{$0.184$} \\[5pt]
				\multirow{1}{*}{$H_a$: $\avgopp^{(\mathrm{SCC},practical,low)} \neq \avgopp^{(\mathrm{SCC},partial,medium)}$} & & & \\[5pt]
				\hline
				\multirow{1}{*}{$H_0$: $\avgopp^{(\mathrm{SCC},complete,low)} = \avgopp^{(\mathrm{SCC},partial,medium)}$} & \multirow{2}{*}{\color{red}$0.000$} & \multirow{2}{*}{\color{green}$0.964$} & \multirow{2}{*}{$0.036$} \\[5pt]
				\multirow{1}{*}{$H_a$: $\avgopp^{(\mathrm{SCC},complete,low)} \neq \avgopp^{(\mathrm{SCC},partial,medium)}$} & & &\\[5pt]
				\hline
				\multirow{1}{*}{$H_0$: $\avgopp^{(\mathrm{SCC},complete,low)} = \avgopp^{(\mathrm{SCC},practical,medium)}$} & \multirow{2}{*}{\color{red}$0.000$} & \multirow{2}{*}{\color{green}$0.750$} & \multirow{2}{*}{$0.250$} \\[5pt]
				\multirow{1}{*}{$H_a$: $\avgopp^{(\mathrm{SCC},complete,low)} \neq \avgopp^{(\mathrm{SCC},practical,medium)}$} & & & \\[5pt]
				\hline
				\multirow{1}{*}{$H_0$: $\avgopp^{(\mathrm{SCC},practical,low)} = \avgopp^{(\mathrm{SCC},partial,high)}$} & \multirow{2}{*}{\color{red}$0.000$} & \multirow{2}{*}{\color{green}$0.787$} & \multirow{2}{*}{$0.213$} \\[5pt]
				\multirow{1}{*}{$H_a$: $\avgopp^{(\mathrm{SCC},practical,low)} \neq \avgopp^{(\mathrm{SCC},partial,high)}$} & & & \\[5pt]
				\hline
				\multirow{1}{*}{$H_0$: $\avgopp^{(\mathrm{SCC},complete,low)} = \avgopp^{(\mathrm{SCC},partial,high)}$} & \multirow{2}{*}{\color{red}$0.000$} & \multirow{2}{*}{\color{green}$0.958$} & \multirow{2}{*}{$0.042$} \\[5pt]
				\multirow{1}{*}{$H_a$: $\avgopp^{(\mathrm{SCC},complete,low)} \neq \avgopp^{(\mathrm{SCC},partial,high)}$} & & & \\[5pt]
				\hline
				\multirow{1}{*}{$H_0$: $\avgopp^{(\mathrm{SCC},practical,medium)} = \avgopp^{(\mathrm{SCC},partial,high)}$} & \multirow{2}{*}{\color{red}$0.000$} & \multirow{2}{*}{\color{green}$0.858$} & \multirow{2}{*}{$0.142$} \\[5pt]
				\multirow{1}{*}{$H_a$: $\avgopp^{(\mathrm{SCC},practical,medium)} \neq \avgopp^{(\mathrm{SCC},partial,high)}$} & & & \\[5pt]
				\hline
				\multirow{1}{*}{$H_0$: $\avgopp^{(\mathrm{SCC},complete,medium)} = \avgopp^{(\mathrm{SCC},partial,high)}$} & \multirow{2}{*}{\color{red}$0.000$} & \multirow{2}{*}{\color{green}$0.955$} & \multirow{2}{*}{$0.045$} \\[5pt]
				\multirow{1}{*}{$H_a$: $\avgopp^{(\mathrm{SCC},complete,medium)} \neq \avgopp^{(\mathrm{SCC},partial,high)}$} & & & \\[5pt]
				\hline
				\multirow{1}{*}{$H_0$: $\avgopp^{(\mathrm{SCC},complete,low)} = \avgopp^{(\mathrm{SCC},practical,high)}$} & \multirow{2}{*}{\color{red}$0.000$} & \multirow{2}{*}{\color{green}$0.785$} & \multirow{2}{*}{$0.215$} \\[5pt]
				\multirow{1}{*}{$H_a$: $\avgopp^{(\mathrm{SCC},complete,low)} \neq \avgopp^{(\mathrm{SCC},practical,high)}$} & & &\\[5pt]
				\hline
				\multirow{1}{*}{$H_0$: $\avgopp^{(\mathrm{SCC},complete,medium)} = \avgopp^{(\mathrm{SCC},practical,high)}$} & \multirow{2}{*}{\color{red}$0.000$} & \multirow{2}{*}{\color{green}$0.782$} & \multirow{2}{*}{$0.218$} \\[5pt]
				\multirow{1}{*}{$H_a$: $\avgopp^{(\mathrm{SCC},complete,medium)} \neq \avgopp^{(\mathrm{SCC},practical,high)}$} & & & \\[5pt]
				\hline
			\end{tabular}
		\end{adjustbox}
	\end{center}
\end{table}

\end{document}